\documentclass{aastex631}

\newcommand\visit{\emph{Visit}}
\newcommand\exposure{\emph{Exposure}}

\accepted{October 16, 2023}

\begin{document}

\title{The Near-Earth Object Surveyor Mission}


\author[0000-0002-7578-3885]{A. K. Mainzer}
\affiliation{University of Arizona, 1629 E University Blvd, Tucson, AZ 85721-0092, USA}

\author[0000-0003-2638-720X]{Joseph R. Masiero}
\affiliation{Caltech/IPAC, Pasadena, CA, USA}

\author[0000-0002-6233-1820]{Paul A. Abell}
\affiliation{NASA Johnson Space Center, 2101 NASA Parkway, Houston TX 77058}

\author[0000-0001-9542-0953]{J. M. Bauer}
\affiliation{Dept. of Astronomy, Univ. of Maryland, College Park, MD}

\author[0000-0002-1804-7814]{William Bottke}
\affiliation{Southwest Research Institute, Boulder, CO, USA}

\author[0000-0002-5901-4875]{Bonnie J. Buratti}
\affiliation{Jet Propulsion Laboratory, California Institute of Technology, Pasadena, CA, USA}

\author[0000-0002-0221-6871]{Sean J. Carey}
\affiliation{IPAC, California Institute of Technology, 1200 E. California Blvd, Pasadena, CA 91125, USA}

\author[0000-0002-5468-1564]{D. Cotto-Figueroa}
\affiliation{Department of Physics and Electronics, University of Puerto Rico at Humacao, Call Box 860, Humacao, PR 00792, USA}

\author{R. M. Cutri}
\affiliation{IPAC, Mail Code 100-22, California Institute of Technology, 1200 E. California Blvd., Pasadena, CA 91125, USA}

\author[0000-0003-1876-9988]{D. Dahlen}
\affiliation{IPAC, Mail Code 100-22, California Institute of Technology, 1200 E. California Blvd., Pasadena, CA 91125, USA}

\author{Peter R. M. Eisenhardt}
\affiliation{Jet Propulsion Laboratory, California Institute of Technology, Pasadena, CA, USA}

\author[0000-0003-1156-9721]{Y. R. Fernandez}
\affiliation{University of Central Florida, Orlando, FL, USA}

\author[0000-0001-6076-8992]{Roberto Furfaro}
\affiliation{University of Arizona, 1629 E University Blvd, Tucson, AZ 85721-0092, USA}

\author[0000-0002-3379-0534]{Tommy Grav}
\affiliation{University of Arizona, 1629 E University Blvd, Tucson, AZ 85721-0092, USA}

\author{T. L. Hoffman}
\affiliation{Jet Propulsion Laboratory, California Institute of Technology, Pasadena, CA, USA}

\author[0009-0002-1596-071X]{Michael S. Kelley}
\affiliation{Planetary Defense Coordination Office, NASA Headquarters, 300 E Street SW, Washington, DC  20546}

\author[0000-0002-4676-2196]{Yoonyoung Kim}
\affiliation{University of Arizona, 1629 E University Blvd, Tucson, AZ 85721-0092, USA}

\author[0000-0003-4269-260X]{J.\ Davy Kirkpatrick}
\affiliation{IPAC, Mail Code 100-22, California Institute of Technology, 1200 E. California Blvd., Pasadena, CA 91125, USA}

\author{Christopher R. Lawler}
\affiliation{Jet Propulsion Laboratory/California Institute of Technology, Pasadena, CA, USA}

\author[0000-0002-7696-0302]{Eva Lilly}
\affiliation{Planetary Science Institute, Tucson AZ, USA}

\author{X.\ Liu}
\affiliation{IPAC, Mail Code 100-22, California Institute of Technology, 1200 E. California Blvd., Pasadena, CA 91125, USA}

\author[0000-0001-7519-1700]{Federico Marocco}
\affiliation{IPAC, California Institute of Technology, 1200 E. California Blvd, Pasadena, CA 91125, USA}

\author[0000-0003-0107-7803]{K. A. Marsh}
\affiliation{IPAC, California Institute of Technology, 1200 E. California Blvd, Pasadena, CA 91125, USA}

\author[0000-0002-8532-9395]{Frank J. Masci}
\affiliation{IPAC, California Institute of Technology, 1200 E. California Blvd, Pasadena, CA 91125, USA}

\author[0000-0003-1969-4324]{Craig W. McMurtry}
\affiliation{Physics and Astronomy Dept., University of Rochester, New York 14627-0171 USA}

\author[0000-0003-3351-5986]{Milad Pourrahmani}
\affiliation{IPAC, California Institute of Technology, 1200 E. California Blvd, Pasadena, CA 91125, USA}

\author{Lennon Reinhart}
\affiliation{Space Dynamics Laboratory, University of Utah, Logan, UT, USA}

\author[0000-0001-5644-8830]{Michael E. Ressler}
\affiliation{Jet Propulsion Laboratory/Caltech, Pasadena, CA, USA}

\author[0000-0001-5766-8819]{Akash Satpathy}
\affiliation{University of Arizona, Tucson, AZ, USA}

\author[0000-0003-1800-8521]{C. A. Schambeau}
\affiliation{Florida Space Institute, University of Central Florida, Orlando, FL, USA; Department of Physics, University of Central Florida, Orlando, FL, USA}

\author[0000-0003-2762-8909]{S. Sonnett}
\affiliation{Planetary Science Institute, Tucson AZ, USA}

\author{Timothy B. Spahr}
\affiliation{NEO Sciences LLC, Marlborough, MA, USA}

\author[0000-0001-7291-0087]{Jason A. Surace}
\affiliation{IPAC, California Institute of Technology, 1200 E. California Blvd, Pasadena, CA 91125, USA}

\author{Mar Vaquero}
\affiliation{Jet Propulsion Laboratory, California Institute of Technology, Pasadena, CA, USA}

\author[0000-0001-5058-1593]{E. L. Wright}
\affiliation{University of California, Los Angeles, Los Angeles, CA, USA}

\author[0000-0002-0515-4813]{Gregory R. Zengilowski}
\affiliation{University of Arizona, Tucson, AZ, USA}

\collaboration{40}{NEO Surveyor Mission Team}

\begin{abstract}

The Near-Earth Object (NEO) Surveyor mission is a NASA observatory designed to discover and characterize near-Earth asteroids  and comets. The mission's primary objective is to find the majority of objects large enough to cause severe regional impact damage ($>$140 m in effective spherical diameter) within its five-year baseline survey. Operating at the Sun-Earth L1 Lagrange point, the mission will survey to within 45 degrees of the Sun in an effort to find the objects in the most Earth-like orbits. The survey cadence is optimized to provide observational arcs long enough to reliably distinguish near-Earth objects from more distant small bodies that cannot pose an impact hazard. Over the course of its survey, NEO Surveyor will discover $\sim$200,000 - 300,000 new NEOs down to sizes as small as $\sim$10 m and thousands of comets, significantly improving our understanding of the probability of an Earth impact over the next century.

\end{abstract}


\section{Introduction} \label{sec:intro}
Asteroid and comets have impacted the Earth for billions of years and will continue to do so. The present-day impact flux is dominated by a supply of asteroidal material that predominantly originates from the main belt between Mars and Jupiter, typically migrating into near-Earth space via thermal drifts that push the objects into gravitational resonances capable of changing orbits on timescales of millions- to tens-of-millions of years \citep{Bottke.2002a, Granvik.2018a}. Once objects evolve into orbits with perihelia less than 1.3 au, they are classified as NEOs.\footnote{https://cneos.jpl.nasa.gov/about/neo\_groups.html} Comets, both long- and short-period, constitute a smaller fraction of NEOs, estimated to be between 5-15\% of the total \citep{Wetherill.1987a, Wetherill.1988a, Bottke.2002a, Fernandez.2005a, DeMeo.2008a, Kim.2014a, Bauer.2017a, Granvik.2018a}. The size distribution is such that material that collides with Earth is dominated by tiny dust particles that produce harmless ``shooting stars"; roughly 100 tons of such material (thought to be primarily cometary in origin) falls on Earth each day \citep{Jenniskens.2015a, Nesvorny.2010a}.\footnote{https://cneos.jpl.nasa.gov/about/target\_earth.html} 

Larger objects impact our planet infrequently. While such impacts are rare, they are capable of causing significant damage locally \citep[e.g. the fireball explosion over Chelyabinsk, Russia in 2013 and the 1908 Tunguska blast;][]{Brown.2013a, Jenniskens.2009a} or globally \citep[e.g. the Chixulub impact 66 Myr ago; ][]{Alvarez.1980a, Renne.2013a}. The impact frequency as a function of impactor size is reasonably well-understood on astronomical or geological timescales, based on studies of the cratering records of the Earth and Moon, as well as telescopic studies of the NEO population. Earth impacts capable of causing global destruction are thought to occur every $\sim$100 million years; impacts capable of causing severe regional destruction might occur every few thousand to tens of thousands of years; and impactors capable of causing damage to a city occur perhaps 0.1 to several times each century \citep{Shoemaker.1979a, Shoemaker.1983a, Rabinowitz.2000a, Stuart.2004a, Mainzer.2011e, Trilling.2017a, Granvik.2018a, Harris.2021a}. Yet to predict what \emph{will} occur on human timescales (over the next century or two) requires discovery of individual objects, with their orbits determined to sufficient precision to assess whether or not an impact is likely over that timescale. 

Much of what we know about the population of NEOs derives from systematic surveys for them, beginning with photographic plate surveys such as the Palomar Planet-Crossing Asteroid Survey \citep{Helin.1979a}. As charged-coupled devices matured and were incorporated into asteroid searches, the pace of NEO discovery increased due to the efforts of projects such as the Spacewatch survey \citep{McMillan.2007a}, the Near-Earth Asteroid Tracking Program \citep{Helin.1997a, Pravdo.1999a}, the Lowell Observatory Near-Earth Object Survey \citep{Koehn.2000a}, and the Lincoln Near-Earth Asteroid Research Program \citep{Stokes.2000a}. Current NASA-supported NEO surveys include the Catalina Sky Survey \citep{Larson.2007a}, PanSTARRS \citep{Chambers.2016a}, the Zwicky Transient Facility \citep{Masci.2019a}, the Near-Earth Object Wide-field Infrared Survey Explorer (NEOWISE) survey \citep{Cutri.2012a, Mainzer.2011a, Mainzer.2014b}, and the Asteroid Terrestrial-impact Last Alert System \citep{Tonry.2018a}. 

As a result of the community's efforts, more than 90\% of the population of 1 km and larger NEAs were believed to have been discovered \citep{Mainzer.2011e, Granvik.2018a} by the 2010-2011 timeframe, fulfilling the ``Spaceguard" goal \citep{Morrison.1992a} of finding the majority of these large asteroids capable of causing global extinction events. The Spaceguard objective did not cover comets.

Following the achievement of the Spaceguard goal, community consensus studies \citep{Stokes.2003a, Stokes.2017a, NAS.2019a} determined that asteroids larger than 140 m in diameter should be the target of the next generation of surveys because they are capable of causing severe regional destruction with economic effects that would be felt globally; a 140 m impactor has the equivalent energy of roughly 200 megatons of TNT. In 2005, the United States Congress passed the George E. Brown, Jr. Near-Earth Object Survey Act requiring that NASA detect more than 90\% of all NEOs larger than 140 m in diameter by the year 2020.\footnote{https://www.congress.gov/bill/109th-congress/house-bill/1022/text} Completing the survey of 140 m and larger NEOs is a key component of the United States' National Near-Earth Object Preparedness Strategy and Action Plan \citep{OSTP.2018a, OSTP.2023a}. At present, perhaps 40\% of all such objects have been discovered with the current suite of surveys \citep{Mainzer.2011e, Granvik.2018a}. Discoveries are currently dominated by the Catalina Sky Survey and PanSTARRS survey, which employ 1-2 m class telescopes operating at visible wavelengths and discover $\sim$2500-3000 new NEOs per year. The average size of newly discovered NEOs is between 30-100 m; the current rate of discovery of larger objects is limited to $\sim$450 new objects per year.\footnote{https://cneos.jpl.nasa.gov/stats/size.html} 

Not all NEOs pose equal hazards to the Earth: in particular, those with orbits that approach within 0.05 au of Earth's orbit have sufficient uncertainty in their ephemerides over the next 80-100 years that impacts beyond that timescale cannot be ruled out \citep{Giorgini.2002a, Ostro.2004a}. Such objects are formally designated potentially hazardous asteroids (PHAs) if they also have absolute magnitude $H<$22 mag \citep[although others have argued for a size-based definition for objects large enough to create a crater, e.g.][]{Mainzer.2014a}. For our purposes in this paper, we include any object with minimum orbit intersection distance (MOID) $<$0.05 au in the definition of PHAs. NEOs are divided into different dynamical classes: Atiras are NEOs with orbits entirely interior to Earth's, with semi-major axes a$<$1.0 and aphelia Q$<$0.983 au. Atens are defined as NEOs with a$<$1 au and Q$>$0.983 AU; Apollos have a$>$1 AU and perihelia distances q$<$1.017 AU. Amors spend all of their time outside the orbit of the Earth, with a$>$1 au and 1.017$<$q$<$1.3 au \citep{Belton.2004a}. Thus, both Atens and Apollos have orbits that cross Earth’s, but Atens tend to have the most circular, Earth-like orbits. Amors generally have the least chance of making close approaches to Earth. It is important to focus on finding PHAs since they have the greatest potential for close Earth approaches.

The challenge of finding 90\% of the \emph{diameter-limited} population of NEAs larger than 140 m is exacerbated by the presence of objects with low-albedo, carbonaceous material among the population. The survey of such objects performed by the Wide-field Infrared Survey Explorer \citep[WISE;][]{Wright.2010a} during the fully cryogenic part of its mission, which detected NEOs based on their 12 $\mu$m thermal emission and was therefore largely insensitive to variations in albedo, found that the NEO albedo distribution remains relatively unchanged as a function of size over the range of diameters probed by WISE \citep[$\sim$200-300m and larger;][]{Mainzer.2011e}. \citet{Wright.2016a} showed that in order to discover 90\% of the 140 m and larger objects, surveys must reach an equivalent absolute magnitude completeness of $H<$23 mag instead of the $H<$22 mag assumed using the approximation that all NEOs have visible geometric albedos of $\sim$0.14. Thus, reaching the 90\% completeness limit specified by the George E. Brown Act requires discovering the dark portion of the NEO population as well as the brighter objects with silicate-dominated surfaces.

Finding PHAs well in advance of any potential close approaches is important because successfully deflecting an impacting object likely requires years to decades of lead time. The Double Asteroid Redirection Test (DART) mission has recently succeeded in altering the orbit of an asteroid using the kinetic impactor technique, following on the success of the Deep Impact \citep{Ahearn.2005a} and Hayabusa2 \citep{Watanabe.2019a} missions. However, such kinetic deflections typically only work on sub-km-sized objects with $\sim$5-50 years of warning, and km-sized bodies would require even more time to deflect them successfully; see Figure 5.5 in \citet{NAS.2010a}. For example, the DART mission was able to impart a total velocity change of 3 mm/sec on its target body, the 160 m diameter Dimorphos (a natural moon of the 765 m NEO Didymos), using a $\sim$600 kg spacecraft \citep{Cheng.2016a}. Assuming a spacecraft of similar mass as DART was sent to deflect an impactor, it would take decades to deflect a 160 m object by the 6400 km radius of the Earth. This time could be cut down by impacting with a more massive spacecraft, or multiple spacecraft, but the time required for a successful kinetic deflection is still likely to be in excess of 20 years. The mass of a larger object scales with the cube of the diameter, which dramatically increases the required deflection time.  Therefore, it is essential to complete the survey of objects large enough to cause at least severe regional damage as quickly as possible. Maximizing warning time minimizes the energy required to deflect an object and allows for the time needed to develop and mature the needed mitigation mission(s) and precursors.

The project completed its Preliminary Design Review in September 2022. NEO Surveyor was confirmed to enter Phase C in November 2022 and is scheduled for launch in September 2027 \citep{Hoffman.2022a}. NEO Surveyor was recommended as the top priority for planetary defense in the 2022 Planetary Decadal Survey \citep{NAS.2022a} and in a previous consensus study paper \citep{NAS.2019a}.

\section{Mission Architecture}

\begin{figure}[ht!]
\plotone{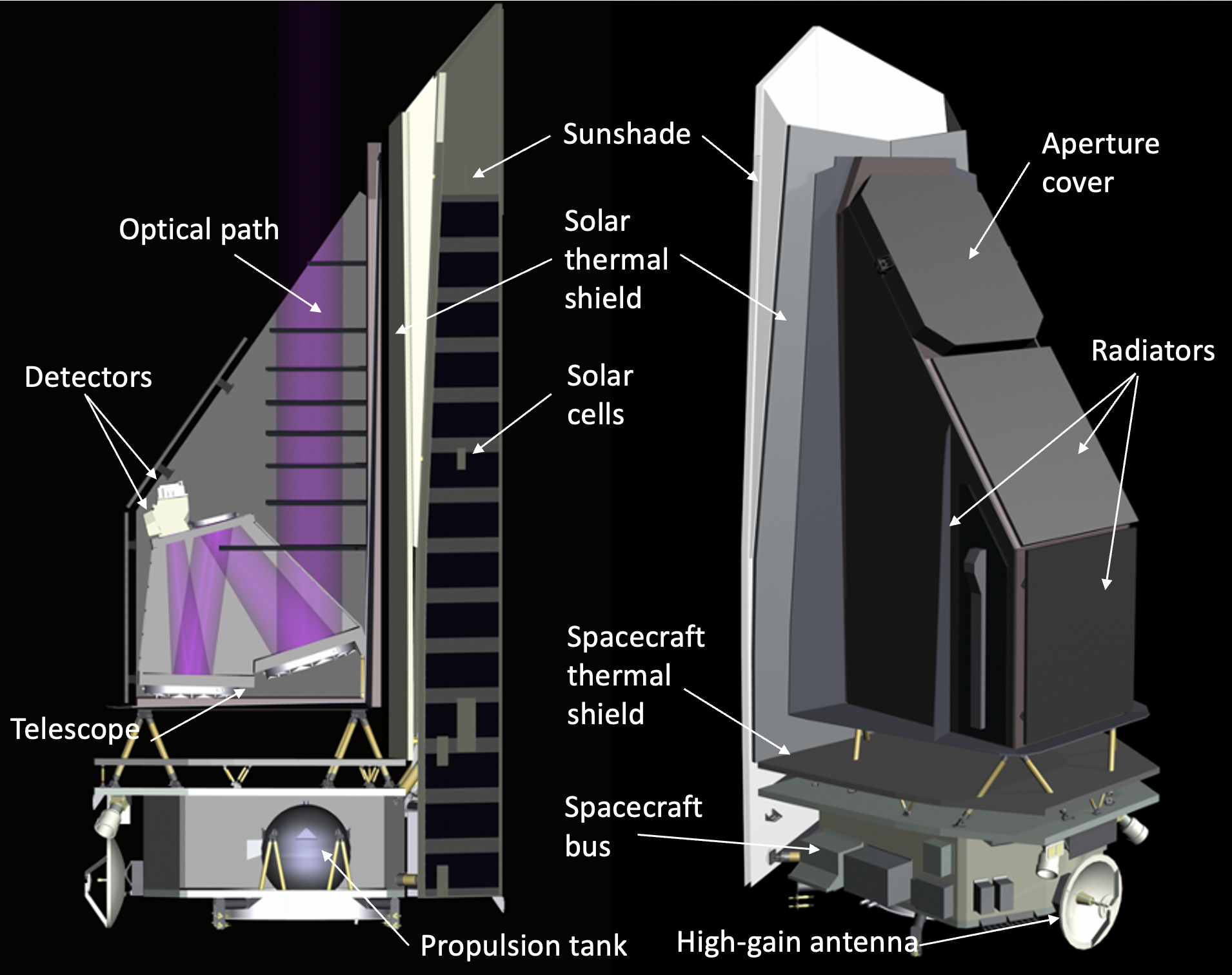}
\caption{The NEO Surveyor mission consists of a single infrared instrument operating at two infrared channels. The instrument is shielded from the Sun by its sunshade, and the black-painted radiator surfaces on the back of the instrument enable the mission to achieve its operating temperatures through passive cooling. The NEO Surveyor mission's 6.15 m sunshade enables it to look down to solar elongations of 45$^{\circ}$ in order to more efficiently detect asteroids in the most circular, Earth-like orbits. The instrument is shown on the right with its one-time deployable aperture cover still attached.
\label{fig:mission_architecture}}
\end{figure}

\begin{figure}[ht!]
\plotone{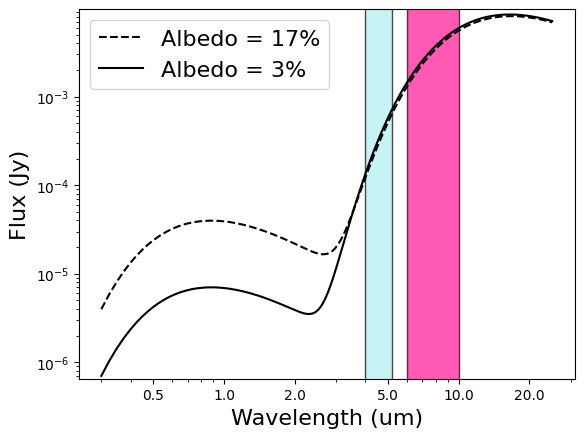}
\caption{The spectral energy distribution for two different 140 m NEOs observed at 90$^{\circ}$ phase angle with visible geometric albedos of 0.03 and 0.17, with the NC1 and NC2 bandpasses overplotted as the cyan and magenta patches, respectively. 
\label{fig:NEO_SEDs}}
\end{figure}

The NEO Surveyor mission is a space telescope dedicated to finding, cataloging, and characterizing the PHAs, including the low albedo objects, as well as long- and short-period comets. It is designed to make significant, rapid progress toward the objective specified by the George E. Brown law. The mission's primary scientific requirement is to find the majority of PHAs larger than 140 m within its five-year baseline mission. The observatory consists of a 50-cm infrared telescope operating at two channels that are dominated by thermal infrared emission for typical NEOs (Figure \ref{fig:mission_architecture}), which typically have effective temperatures between 200-300 K throughout the majority of their orbits (Figure \ref{fig:NEO_SEDs}). Channel NC1, spanning 4-5.2 $\mu$m, is designed to detect background stars for obtaining astrometric registration and calibration as well as for improving constraints on an object's effective temperature. Channel NC2 spans 6-10 $\mu$m in an effort to maximize sensitivity to typical NEO thermal emission. 

The mission's driving Level 1 requirement is to find 2/3 of PHAs larger than 140 m within its 5-year baseline mission. The mission has a goal of operating for 12 years, which would give the best chance of achieving 90\% completeness on such objects, thus fulfilling the George E. Brown law. NEO Surveyor is also required to constrain the impact frequency of smaller NEOs and the total population of comets. In addition, the Observatory is required to provide the capability to stop and collect additional follow up observations of targets of interest, should an object be discovered with a particularly large chance of impact. The 3-axis stabilized spacecraft is capable of stopping and integrating on any individual target of interest in the instrument operable zone. The mission's concept of operations is designed to maximize discovery of PHAs and comets. As described below, after operating for $\sim$10-12 years, the Observatory has a reasonable chance of reaching 90\% completeness for PHAs $>$140 m, representing $\sim$200,000-300,000 NEOs down to small sizes.

\begin{figure}[ht!]
\plotone{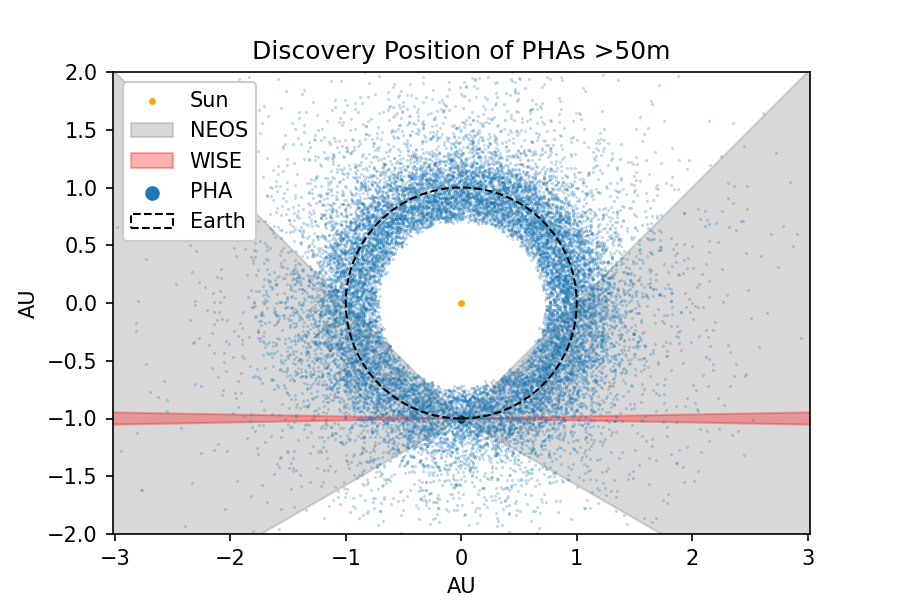}
\caption{The NEO Surveyor field of regard (gray cones) is shown compared to that of NEOWISE (pink cones) and the orbit of the Earth (dashed line). The simulated population of PHAs larger than 50 m at the time of discovery is shown as blue dots. The NEOWISE field of regard is limited by its relatively short sunshade and low-Earth orbit, which prohibits it from reaching the near-Sun regions on the sky where NEOs are more likely to be found. NEO Surveyor's sunshade is tall enough to enable it to observe down to 45$^{\circ}$ from the Sun, subtending a larger portion of the PHAs.
\label{fig:field_of_regard}}
\end{figure}

By entering into a halo orbit around the Sun-Earth L1 Lagrange point (SEL1), the NEO Surveyor mission is able to continuously view a large segment of the space around Earth's orbit, particularly the near-Sun regions that are most difficult to observe from the ground (Figure \ref{fig:field_of_regard}). The SEL1 halo orbit is preferred over the L2 Lagrange point because the Moon does not impinge on the field of regard as significantly, and the spacecraft's communications antenna does not have to be mounted on the solar panel/sunshade, simplifying the design. Previous studies \citep{Mainzer.2015a, Grav.2016a} indicated that an SEL1 orbit yielded comparable performance to a circulating interior-to-Earth or Venus-trailing orbits, even assuming that no efficiency losses were incurred from the several orders-of-magnitude drop in data downlink rate that would be a consequence of such orbits. 

At SEL1, the Earth and Moon are sufficiently distant that their thermal radiation onto the spacecraft can be managed purely passively. Combined with the other thermal protections, this design allows the NEO Surveyor instrument performance to be limited by the natural astrophysical background in the NC2 channel \citep[predominantly zodiacal dust emission;][]{Wright.1998a, Leinert.1998a}, with the telescope required to remain below 57 K. The NC2 detectors must be maintained below 40 K \citep{McMurtry.2013a, Dorn.2016a, Zengilowski.2021a, Zengilowski.2022a} to ensure that their dark currents do not dominate over the minimum expected zodiacal background signal over the field of regard. The detectors are nonlinearly sensitive to temperature, and the temperature requirements are designed to limit heat input into the arrays. The multi-stage radiator system includes shields to intercept heat from the solar panel and spacecraft. The observatory's sunshade blocks sunlight and carries the solar panels. A shield underneath the instrument intercepts heat emitted from the $\sim$300 K spacecraft bus, and three radiators coated with high-emissivity black paint affixed to the back of the instrument reject heat from the thermal shields, readout electronics, and detectors. The passive thermal system does not require expendable cryogens or cryocoolers (Figure \ref{fig:mission_architecture}). The thermal system employs industry-standard thermal margins and design practices to stabilize the telescope, baffles, and focal plane temperatures \citep{Hoffman.2022a} and derives heritage from the Spitzer Space Telescope \citep{Finley.2004a, Finley2005a, Lawrence.2004a}. 

From SEL1, it is possible to downlink at 150 Mbps using NASA's 35-m Deep Space Network dishes. This high-rate data link allows for the transmission of the individual 30-second \textit{Exposures} collected by the instrument in both NC1 and NC2 channels simultaneously that are used to construct each 180-second observation.  

The telescope optical system consists of a three-mirror anastigmat design that provides a field of view of 1.68$^{\circ}$ x 7.08$^{\circ}$ with no central obscuration, thus maximizing sensitivity and eliminating diffraction spikes due to the secondary mirror support structure. There is no focus mechanism; focus is set through a series of  measurements made on the ground at the operating temperature that account for the deformation of the optical system as it cools down. The same field of view is imaged simultaneously by the two channels using a germanium beamsplitter, with NC1 being reflected and NC2 transmitted through the beamsplitter. Both channels utilize transmission filters to define their bandpasses.

Each focal plane consists of a 4x1 array of 2048x2048-pixel HgCdTe detectors that are closely packed together to minimize the gaps between individual arrays that can cause loss of moving object detections and thus break links between detections. The gap spacing between arrays is 2.4 mm for both channels, with a plate scale of $\sim$3 arcsec per 18 $\mu$m pixel. 

The instrument can safely point anywhere from 45-125$^{\circ}$ in elongation from the Sun including both ecliptic poles; this is defined as the instrument operable zone. The field of regard of the survey is a subset of this area, spanning 45-120$^\circ$ in longitudinal distance from the Sun and stopping at $\pm$40$^{\circ}$ ecliptic latitude (Figure \ref{fig:field_of_regard}). Calibration fields are measured at the ecliptic poles once per week, since a large set of well-characterized calibration targets are found in these regions \citep{Reach.2005a, Jarrett.2011a, Carey.2012a}. 

Calibration for NEO Surveyor focuses on two high level objectives, astrometric accuracy of $<$0.5 arcseconds (1 sigma rms) for sources of SNR$\geq$20 and photometric accuracy of better than 7\% for bright (non photon-noise limited), unsaturated point sources. These requirements  help to ensure that the survey produces data of sufficient quality to accurately measure orbits and diameters of the population of near-Earth objects greater than 140 meters in size.  Using photometric calibration methods demonstrated with Spitzer and WISE and astrometric calibration based on the Gaia astrometric frame of reference, the absolute photometric calibration is expected to be better than 4\% and the astrometric accuracy for SNR$\geq$20 sources to be better than 0.3 arcsec per detection.

The sensitivity of the Observatory varies over the sky due to the wide range of solar longitudes covered by the field of regard. In both NC1 and NC2, the sensitivity is dominated by the natural zodiacal background, which varies by roughly a factor of 20 and 10 in NC1 and NC2, respectively. From a zodiacal background model based on DIRBE measurements, we estimate the zodiacal background to range from $\sim$220-4200 nWm$^{-2}$sr$^{-1}$ in NC1 and from $\sim$1700-22,000 nWm$^{-2}$sr$^{-1}$ in NC2 across the NEO Surveyor field of regard \citep{Wright.1998a, Leinert.1998a}. Since we plan to extract point sources at a signal-to-noise ratio of 5 or greater, we quantify sensitivity as five times the noise equivalent spectral irradiance (denoted $NESI_{5}$). The sensitivity calculation takes into account the sampling of the NEO Surveyor instrument's point spread function, the integration times in both channels, the telescope entrance pupil area, the thermal self-emission, stray light, dark current, and other parameters. The $NESI_{5}$ is computed across the full field of regard and is required to be within 65-120 $\mu$Jy in the NC1 channel and 110-280 $\mu$Jy in the NC2 channel.

\section{Concept of Operations}
The mission's science operations consist of a highly repetitive survey pattern that is optimized for potentially hazardous object detection and is continuously executed throughout the baseline mission. Each day, the survey pattern is executed continuously, breaking only to downlink for a total of 2.25 hours/day, to perform calibrations, and occasionally to perform station-keeping and momentum management maneuvers. 

\begin{figure}[ht!]
\plotone{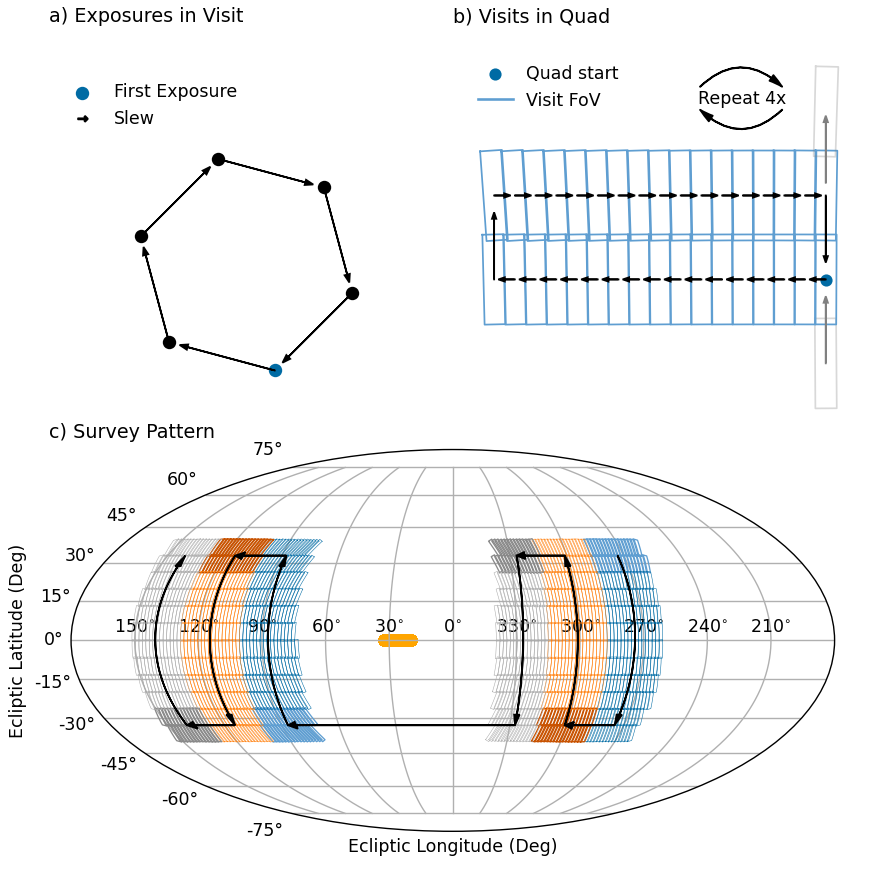}
\caption{Overview of the survey pattern. (a) Each NEO Surveyor \visit\ consists of six \emph{Exposures} collected in a hexagonal dither pattern.  (b) \emph{Visits} are repeated in a \emph{Loop} four times to make a \emph{Quad}. (c) \emph{Quads} are then tiled into \emph{Stacks}, then \emph{Sides} of the Sun (Orange line). This figure shows two full \emph{Sides} during the survey, made up of three \emph{Stacks} each; darker colors indicate the first \emph{Quad} in each stack. This pattern is then repeated.
\label{fig:survey_pattern}}
\end{figure}

The essential unit of the survey is the \visit, which consists of a set of six dithered individual \emph{Exposures}, each 30 sec long and collected in a roughly hexagonal pattern with a $\sim$10 arcsec step length (Figure \ref{fig:survey_pattern}). The dither step size and number of steps is a configurable table that can be adjusted in flight if needed. The six \emph{Exposures} are co-added together to form each \visit, and in order to measure the photometry of individual sources, the stack of \emph{Exposures} is fit simultaneously as appropriate for data taken with an undersampled system. A total of three minutes are allocated per \visit, with an integration time of $\sim$145 s in both bands. All individual \emph{Exposures} are downlinked each day, facilitating improved artifact and noise rejection, searches for very-fast-moving NEOs, and for so-called ``precovery" detections of moving objects discovered at a later time and subsequently recovered in the NEO Surveyor images.

Within each \exposure, the NC1 and NC2 arrays each have their own unique clocking and readout schemes that are optimized to the astrophysical background level anticipated in each channel (including zodiacal light emission). NC1 collects 18 samples over the course of 26.6 sec in a ``sample-up-the-ramp" mode with a 25.1 sec total net integration time for the ramp; the slope that is fitted to the samples is downlinked, forming a single \exposure. Saturation detection is provided by the \citet{Zemcov.2016a} algorithm that fits the slope to non-saturated portions of the ramp. For NC2, eight individual correlated double sample (CDS) pairs of samples are collected in 27 sec (using 2.90 sec integration time per CDS pair). The eight CDS pairs are co-added using a single-sided outlier rejection algorithm to provide resilience against cosmic rays, forming the \exposure\ for that channel which is downlinked. A single dither step is executed after each \exposure. For one \exposure\ in each dither sequence, the initial NC2 `pedestal' frame (nominally taken 50 msec after detector reset) of the first CDS pair is also downlinked to provide a constraint on the photometry of sources that saturate within 2.90 sec.

Each patch of sky in the field of regard is covered a total of four times over $\sim$6-9 hours. The Observatory steps in the short dimension of the 4x1 detector array to maximize efficiency for a total of 17 \emph{Visits} before stepping up or down in ecliptic latitude and moving across to form a \emph{Loop}. This looping pattern is repeated until four \emph{Visits} are collected at each pointing on the sky (four \emph{Loops} are denoted a \emph{Quad}). The pairs of positions and times collected from detections harvested from \emph{Visits} in each \emph{Quad} are linked together to form a ``tracklet". Next, the survey steps  in ecliptic latitude collecting successive \emph{Quads} until the full range from -40$^{\circ}$ to +40$^{\circ}$ is covered. This set of \emph{Quads} forms a \emph{Stack}; two additional \emph{Stacks} are collected before the Observatory flips over to the opposite side of the sky. Each set of three \emph{Stacks} is called a \emph{Side} (see Figure \ref{fig:survey_pattern}). In the event that spurious sources interfere with reliable linking of legitimate detections of small bodies, it is possible to increase the number of \emph{Loops} performed to five with minimal impact to survey completeness for NEOs $>$140 m. All data processing to reduce the data, including extracting sources and linking tracklets, is performed on the ground.

The detectors collect photons for 56.7\% of the total time, with camera readout overhead taking 11.7\%, slews between dither steps taking 4.9\%, slews between \emph{Visits} taking 15.7\%, and all other activities (including daily downlinks, momentum management, and calibration pointings) taking 10.8\% of the total time. Over the course of the 5-year baseline mission, roughly 630,000 \emph{Visits} are collected. 

NEO Surveyor will also be capable of executing targeted follow-up observations (TFOs) in order to obtain more information on an object of special interest, including refined astrometry and improved photometric characterization. The TFOs will use the same hexagonal pattern for each \visit, as in the nominal survey mode, but the \visit~ pattern will be repeated on a single field for much longer to obtain up to $\sim$20x more exposures relative to a nominal survey \visit. To avoid impacting the overall time available for the survey, the use of TFOs is minimized, since each TFO expends propellant reserves and observing time at the expense of survey progress. Based on the current numbers of objects on the Center for Near Earth Object Studies Sentry table of virtual impactors\footnote{https://cneos.jpl.nasa.gov/sentry/vi.html}, TFOs are not anticipated to take more than 1\% of total available survey time.

\section{Survey Data System}
Data processing and archiving for the NEO Surveyor mission is carried out by the NEO Surveyor Survey Data System (NSDS) that is developed and operated at IPAC/Caltech.  The NSDS is a highly automated, high-throughput hardware, software, and operations system that is optimized to identify candidate moving objects. Elements of the NSDS are based closely on the data systems developed and operated for WISE/NEOWISE and the Zwicky Transient Facility (Cutri et al. 2012; Masci et al. 2019).

The NSDS will deliver moving object tracklets to the IAU Minor Planet Center (MPC) approximately two to three times daily, typically following processing and quality assurance of a \emph{Quad}. Tracklets are reported on average within 72 hours following the observation time of the final detection of the last moving object in a set of delivered tracklets. Included in the 72 hours is the fact that downlinks are conducted once per 24 hours; the $\sim$2 hours needed to downlink; the time needed to transmit the data from ground stations to IPAC; and the time needed to process the data to the point of extracting tracklets. The calibrated \exposure\ and \visit\ images and extracted source data will be publicly released semi-annually via the NASA/IPAC Infrared Science Archive (IRSA).\footnote{\textit{https://irsa.ipac.caltech.edu/}} Along with positions and times, estimated visual magnitudes will be delivered based on an average albedo, as is currently being provided by the NEOWISE mission. These estimated visual magnitudes, while having large uncertainties, are nonetheless intended to provide a guide for observers performing follow up observations using visible light telescopes.

The NSDS is comprised of a series of pipelines that convert raw NEO Surveyor data and telemetry into the mission’s photometrically and astrometrically calibrated data products.  The high level NSDS pipelines and Subsystems are as follows:
\begin{itemize}

\item The {\it Ingest/Raw Image Preparation Subsystem} receives raw image data packets, engineering and housekeeping telemetry, spacecraft navigation data, and constructs raw image products with metadata required for downstream processing.

\item The {\it Exposure Subsystem} first removes instrumental signatures from the raw detector images, masks bad pixels, derives astrometric calibration solutions with respect to reference sources drawn from {\it Gaia}, gathers photometric calibration solutions, and then attaches these to the image metadata. It then detects sources on the calibrated images on both bands simultaneously and performs profile-fitting to measure source photometry in each NEO Surveyor band along with positions in the International Celestial Reference System. Sources identified as instrumental artifacts or contaminated by artifacts are tagged in the output source list.  

\item The {\it Visit Pipeline} combines (by co-adding) the six dithered calibrated \emph{Exposures} in each \visit\ with pixel-outlier masking included. This suppresses transient artifacts such as radiation hits and noisy pixels. This pipeline then performs source detection on both bands simultaneously. Profile-fitting and aperture photometry are then performed on each detection on the deeper combined set of six images in each band. Astrometric positions are also solved for during profile-fitting. As done in the Exposure Pipeline, sources identified as instrumental artifacts or contaminated by artifacts are tagged in the output source list.  

\item The {\it Differencing Pipeline} first performs image differencing by subtracting a sky reference image from both a \visit\ image (co-add) and all overlapping \exposure\ (detector) images therein. Reference images are constructed for each target \visit\ by co-adding \exposure\ images in all other \emph{Visits} that overlap the target \visit\ on a \emph{Quad} pointing. The target \emph{Visit's} \emph{Exposures} are omitted from the reference image co-add so as to not suppress moving-object candidates identified in the target \visit. Sources are then detected on the combined set of difference images per \visit\ on both bands simultaneously. Profile-fitting and aperture photometry are then performed with astrometric positions derived from the profile-fit solutions. The detection list is then filtered to remove spurious sources using a machine-learned (ML) classifier. This classifier uses a combination of a convolutional neural network applied to image data and a transformer-based model applied to extracted source features. These are then fed to a multilayer perceptron to infer the class of the detected source. This filtering results in a ``purified" list of candidate moving objects and includes objects resolved with respect to the point spread function (PSF). The latter constitute probable comets. This list is now ready for tracklet generation (see below).

\item The {\it Moving Object Detection Pipeline (MODP)} first constructs tracklets by linking sources across the purified detection lists for every \visit\ in a \emph{Quad} generated by the Differencing Pipeline (see above). This linking is performed by the NEO Surveyor Moving Object Detection Engine (NMODE), a variant of which is being used for ZTF \citep{Masci.2019a}. This first pass finds the majority of solar system objects, and predominately main belt asteroids (MBAs) within its restricted velocity-search thresholds (0.008 - 8.000$^{\circ}$/day). Following filtering of tracklet detections from this first pass, a ``slow NEO" search is performed by running NMODE with loosened velocity tolerances. Following this, a search for ``fast NEOs" (primarily small objects very close to the spacecraft) is performed using velocity estimates provided by profile-fitting from the Differencing Pipeline. These velocities are used to construct synthetic detections (time-tagged positions) within individual \emph{Visits} and then fed to NMODE to construct additional tracklets using further relaxed velocity-match thresholds. Tracklet quality metrics and ``moving co-add" images (in the co-moving frame of each candidate object), along with ``collapsed photometry" metrics are then computed. The list of tracklets, accompanying quality metrics, moving co-add images, and ML-based scores for every detection from the Differencing Pipeline are then fed into the Automated Tracklet Classification Subsystem (ATC). This subsystem attaches an overall reliability score to each tracklet as well as a score to indicate if it is associated with a probable comet based on extendedness metrics derived from moving co-adds and other contextual metadata.

\item The {\it Survey Data Quality Assurance (SDQA) Subsystem} ingests QA metrics, images, tracklet metadata and scores assigned by the automated classifiers from all processing steps upstream for a \emph{Quad}. These products, along with a subset of tracklets, are human reviewed and manually scored. Any updates to the automated quality scores (for either tracklets, detections therein, or both) are fed back to the automated-classifier training frameworks to improve their performance. A list of tracklets with associated metadata, including associations to known objects and reliability scores from both automated classification and human vetting, is delivered to the MPC. The SDQA Subsystem also trends select quality metrics as the survey proceeds. Key metrics are included in reports and used to monitor survey performance.
  
\item The {\it Archive and Database Preparation Pipeline} collects the location of all image products, source lists, and metrics from the Pipeline Operations system, along with quality scores following human vetting and prepares these into ``load scripts" for ingestion into the NASA/IPAC Infrared Science Archive (IRSA).\footnote{https:irsa.ipac.caltech.edu} Source lists and \visit\ images are delivered and made publicly available through IRSA every six months. 

\item The {\it Static Sky Atlas Image Pipeline} is executed every 12 months to construct deep co-adds mosaicked on $\sim$14$^{\circ}$ x 14$^{\circ}$ sky footprints using all good quality survey \emph{Exposures} processed during the preceding 12 months. These co-adds will have an average depth of $\sim$160 \emph{Exposures} or $\sim$2.8 magnitudes deeper than a single \exposure\ image. Atlas Images are publicly released through IRSA every 12 months.

\item The {\it Image Simulation Subsystem} uses the current best estimates for detector and telescope performance along with catalogs of solar system and astrophysical sources to construct high-fidelity simulations of the image data that will be collected by NEO Surveyor. These simulated images are used to support development, testing, design decisions, explore expected survey performance, and verify key functionality prior to launch.

\end{itemize}

\section{Performance Modeling }
To derive the requirements for the NEO Surveyor mission and to model its performance, a simulator was developed that is capable of replicating the properties of the Observatory, its survey cadence and operational characteristics, and the capabilities of the ground software used to extract detections of moving object candidates and link them together. The survey simulation includes a model representation of the solar system's small bodies, including the targets of the NEO Surveyor mission, near-Earth asteroids and comets, as well as the dominant source of moving background confusion, the main belt asteroids. This model, known as the Reference Small Body Population Model (RSBPM), is being used throughout the mission as a stable ``yardstick" for assessing survey completeness for the NEOs, comets, and their sub-populations. 

The steps of the survey simulation are as follows: \begin{enumerate}
    \item Create a reference population model of target small body populations (NEOs and comets) and background populations (main belt asteroids). 
    \item Generate a list of survey fields, taking into account the required number and temporal spacing of detections for orbit determination, the Observatory's keep-out zones, and requirements for non-survey activities such as daily downlinks and momentum management. 
    \item Propagate the ephemerides of each object to the time of each \visit. 
    \item Evaluate whether the object falls within the Observatory's field of view. 
    \item Evaluate which objects in the field of view are bright enough to be detected above the SNR threshold (SNR$>$5). 
    \item Evaluate whether an individual object has accumulated the minimum number of detections to successfully link the detections together reliably. 
    \item Assess whether the detections fall within the limits for apparent rates of motion (0.008 - 8.000$^{\circ}$/day).
    \item Tabulate numbers of detected objects and assess survey completeness as a function of size and subpopulation.
\end{enumerate}   

The sections below describe these steps in greater detail.

\subsection{Reference Population Model Generation}
The reference population model contains both NEOs and background objects. Although NEO Surveyor's Level 1 requirements focus on the detection of NEAs and comets, a robust model of the background population is necessary to ensure that NEOs can be reliably distinguished from more distant small bodies using the mission's planned survey cadence. At the sensitivity depth of NEO Surveyor, MBAs typically outnumber NEOs by a factor of 1000 in \textit{Visits} taken near the ecliptic plane.  In addition to being used to predict the survey performance, the RSBPM is also used to generate the synthetic solar system object detections included in the NSDS simulated images.

\subsubsection{NEO Model}
The first step in generating the NEO population is to create diameters for all objects in the model. Diameters were generated using the inverse transform method to randomly sample a cumulative size distribution. Following \citet{Stokes.2017a}, the cumulative size distribution was taken to have the functional form of a triple power law with
    $N > D^{-\alpha}$,
where N is the cumulative number of objects, D is the diameter, and $\alpha$ is the slope of the power law. A cumulative slope of –2.75 was assumed for diameters D$>$1.5 km, –1.64 for objects with 70 m$<$D$<$1.5 km, and –3.2 for D$<$70 m. To be conservative, the cumulative size distribution was assumed to be identical for all dynamical groups of NEOs (Atiras, Atens, Apollos, and Amors), although there is some evidence that the slopes may vary within these groups \citep{Mainzer.2012b}. The total number of objects is 993$\pm$38 NEAs at 1 km diameter based on \citet{Mainzer.2011e} and \citet{Granvik.2018a}; the above cumulative size distribution results in a total of 25,000 asteroids larger than 140 m being generated.

To determine the visible geometric albedos for the synthetic objects, we used NEOWISE data as the basis of the model. The sample of NEOs detected by the NEOWISE mission's automated detection software, the WISE Moving Object Processing System \citep[WMOPS;][]{Mainzer.2011a}, represents our best understanding of the NEO albedo distribution. This is because WMOPS detects objects based solely on their thermal infrared fluxes, either at 12 $\mu$m during the phases of the mission when that channel was operational \citep{Mainzer.2011c, Mainzer.2012a}, or at 5 $\mu$m during the remaining $\sim$10 years after the survey to date \citep{Mainzer.2014a, Mainzer.2019a, Nugent.2015a, Nugent.2016a, Masiero.2017a}. By selecting objects based on their thermal infrared flux instead of their reflected sunlight, the NEOWISE sample is relatively unbiased with respect to albedo. To date, $\sim$1380 NEOs have been detected by WMOPS, giving us insight into the relative fractions of low albedo vs. high albedo NEOs for Atens, Apollos, and Amors.  

\begin{figure}[ht!]
\plotone{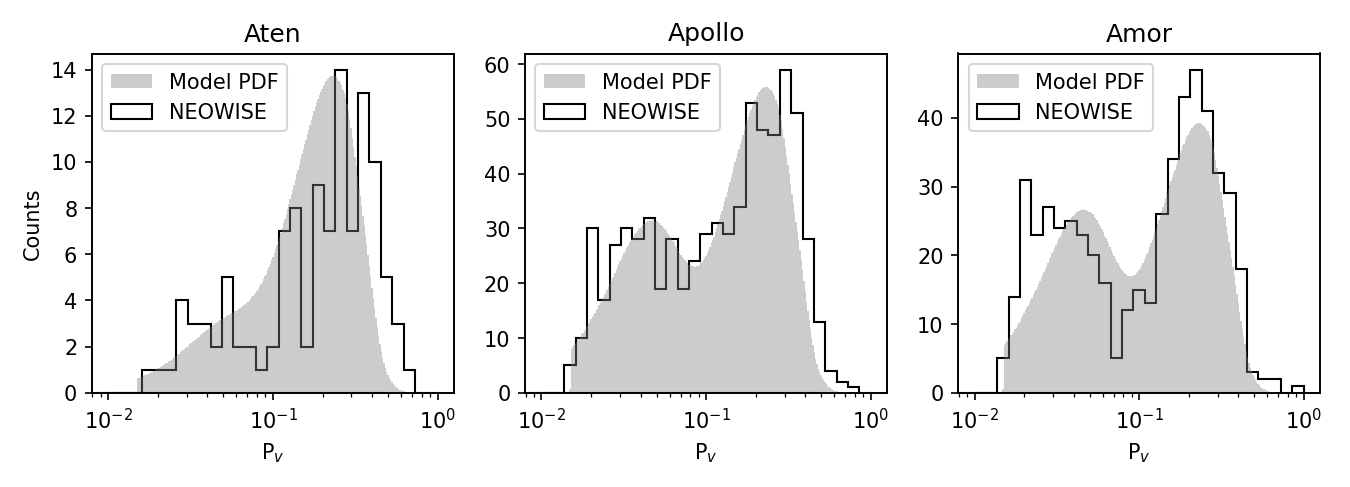}
\caption{Probability distribution functions (solid gray areas) for the Visible geometric albedo distributions $p_{V}$ of the model NEAs compared to the measurements for Atens, Apollos, and Amors detected by the NEOWISE based on their thermal fluxes.
\label{fig:neo_subgroup_albedo_dist}}
\end{figure}

The NEOWISE-detected NEO albedo distribution consists of a group of dark objects with a peak of $\sim$0.03 likely to have carbonaceous compositions and a group of bright objects with a peak of 0.17, most probably associated with stony or metallic compositions. \citet{Wright.2016a} modeled this as a double Rayleigh distribution with peaks centered at 0.03 and 0.17. The fraction of bright to dark objects for each subgroup (Atira, Aten, Apollo, Amor) was used to randomly select whether a given object belonged to the bright or dark group. Since very little is known about the albedo distribution of the Atiras, the fraction of bright to dark Atiras was assumed to be identical to that of the WMOPS-selected Atens. Once a particular object was determined to belong to the bright or dark group, the albedo was then randomly selected using the inverse transform of the cumulative distribution of the appropriate Rayleigh function. Figure \ref{fig:neo_subgroup_albedo_dist} shows the probability distribution functions (PDFs) that were used to generate the albedo distributions of the model compared to the WMOPS-selected albedo distributions for the various NEA dynamical classes. The PDFs are not a perfect match to the individual populations' distributions, partially because they are limited to fall within 0.015 $<p_{V}<$0.60 and partially because NEO albedos depend strongly on absolute $H$ magnitudes, which can have large uncertainties due to the wide range of phase angles at which NEOs are frequently observed. Nonetheless, we tested the simulation with a number of variations of the PDFs, and there was no difference within the statistical uncertainty in the survey performance. 

\begin{figure}[ht!]
\plotone{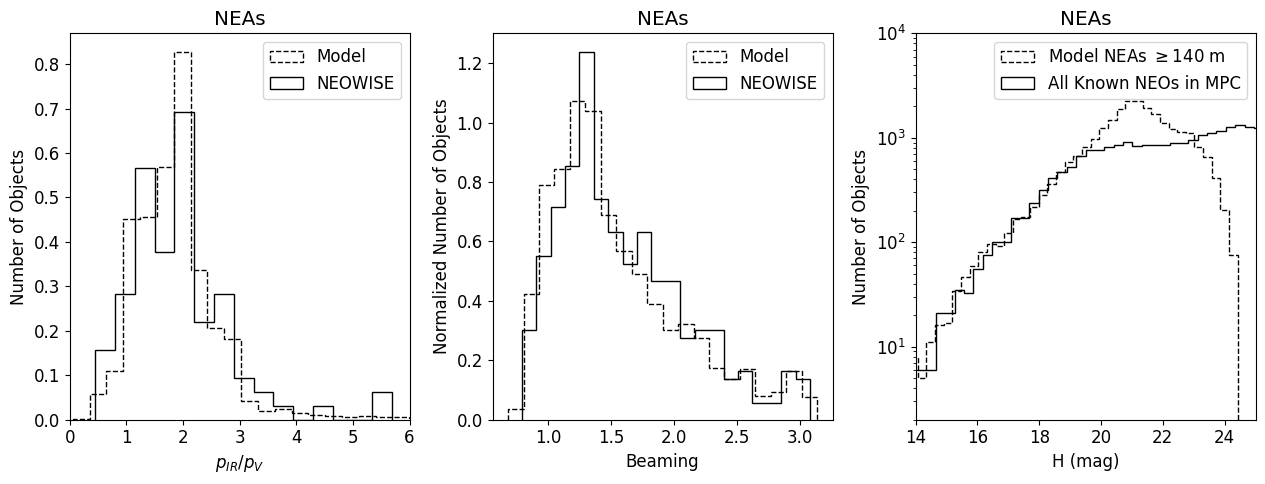}
\caption{(a) Fraction of $p_{IR} / p_{V}$ for the RSBPM model compared to the $\sim$90 NEOWISE-detected NEOs for which $p_{IR}$ could be measured in \citet{Mainzer.2011e}. (b) Distribution of beaming parameters ($\eta$) for the RSBPM model compared to the $\sim$317 NEOWISE-detected NEOs for which $\eta$ could be measured. (c) Distribution of absolute visible magnitudes for the RSBPM model containing only NEAs $\geq$140 m compared to that of the known NEOs to date; the distributions match well until the point at which the population becomes significantly less observationally complete.
\label{fig:neo_phys_prop}}
\end{figure}

The NEO Surveyor NC1 channel spans 4.0-5.2 $\mu$m. For most NEOs over the majority of their orbits, this channel will typically be dominated by thermal emission, but it is important to compute the fraction of reflected sunlight that contributes to the measured flux. Previous studies with NEOWISE were able to fit the albedo at 3.6 $\mu$m \citep{Mainzer.2011c, Masiero.2014a}, but only for populations of more distant objects was the albedo at 4.6 $\mu$m able to be determined separately \citep{Grav.2012a, Grav.2012b}. Here, we assumed that $p_{IR} = p_{3.4 \mu m} = p_{4.6 \mu m}$. The albedo in channel NC1 ($p_{IR}$) was generated by selecting a value for the ratio of $p_{IR}/p_{V}$ using a Gaussian kernel density estimator (KDE) with a PDF drawn from NEOs with fitted $p_{3.6 \mu m}$ values (those with fit code ``DVBI'') found in \citet{Mainzer.2019a}. Fit codes are the values used in \citet{Mainzer.2019a} to denote which parameters were fitted versus assumed in thermal modeling performed on NEOWISE-detected objects. Infrared albedo $p_{IR}$ was limited to 0.015$<p_{IR}<$0.70 (Figure \ref{fig:neo_phys_prop}a).

To facilitate use of the near-Earth asteroid thermal model \citep[NEATM;][]{Harris.1998a} to compute RSBPM small body fluxes, we generated the beaming parameter $\eta$ used by NEATM for each object. Beaming values were similarly selected using the KDE with the PDF drawn from the fitted beaming values in \citet[][fit code ``B'']{Mainzer.2019a} (Figure \ref{fig:neo_phys_prop}b). 

\begin{figure}[ht!]
\plotone{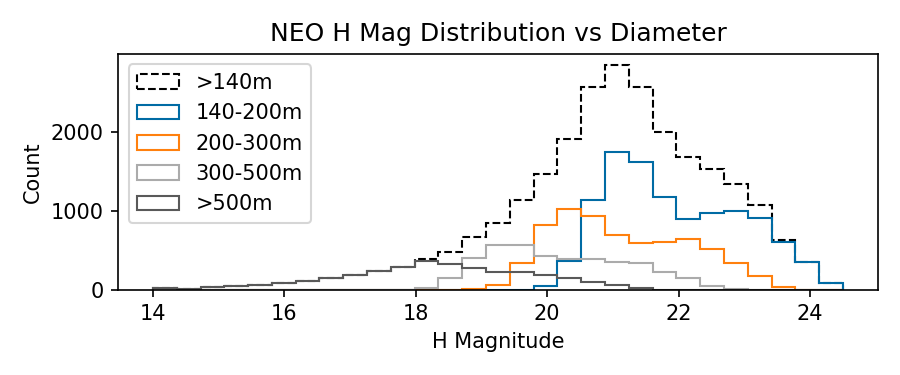}
\caption{The distribution of $H$ magnitudes for NEOs is shown for various diameter bins, illustrating that the canonical assumption that NEAs with $H<$22 mag are all larger than 140 m in diameter is incorrect; many NEOs larger than 140 m have $H>$22 mag.
\label{fig:neo_h_dist}}
\end{figure}

Absolute magnitude $H$ was calculated using the previously generated diameters and $p_{V}$ with the relationship  \begin{equation}
    D (km) = (1329 / \sqrt{p_{V}})  10^{\frac{-H}{5}}
\end{equation} \citep{Fowler.1992a}. Figure \ref{fig:neo_phys_prop}c shows the distribution of $H$ magnitudes in the RSBPM model for NEAs larger than 140 m compared to those of all known NEOs that have been discovered to date (roughly 30,000 objects). Figure \ref{fig:neo_h_dist} shows the distributions of $H$ magnitudes broken down by diameter bins.

\begin{figure}[ht!]
\plotone{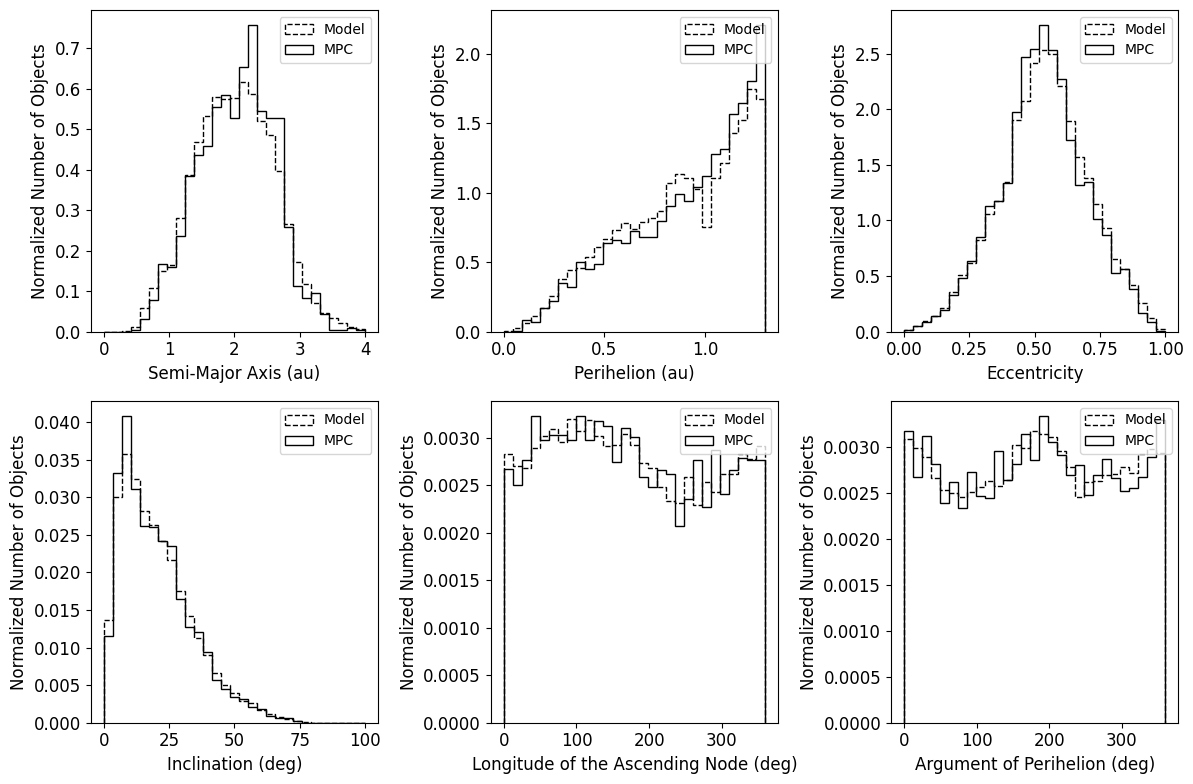}
\caption{Orbital properties of the model NEOs were derived from the orbital elements of all known objects with $H<$20 mag.
\label{fig:neo_orb_props}}
\end{figure}

Orbital elements from the Minor Planet Center's NEO catalog with $H<$20 mag form the basis of the probability distribution functions sampled by a Gaussian KDE to obtain the orbital elements for the synthetic NEOs. This method preserves the observed correlations between orbital elements. At present, more than 90\% of NEOs larger than $\sim$1 km are thought to have been discovered \citep{Mainzer.2011c, Granvik.2018a}, corresponding to an object with $H$=17.75 mag for a visible geometric albedo of 0.14. While the fraction of known NEOs with $H<$20 mag is not observationally complete, this limit ensures an adequate number of objects from which to sample orbital elements and results in a reasonable match to the orbital elements of the observed population (Figure \ref{fig:neo_orb_props}). The relative numbers of Atiras, Atens, Apollos, and Amors were chosen to be 1.4\%, 3.5\%, 55.1\%, and 40.0\% based on the fractions given in \citet{Mainzer.2011e} and \citet{Granvik.2018a}. 

The number of unique representations of each population is determined by the desired uncertainty to be achieved for survey completeness. We can compute the number of objects required in the population using the binomial population confidence interval\footnote{https://en.wikipedia.org/wiki/Binomial\_proportion\_confidence\_interval}, which computes the probability of success from a set of $n$ success-failure trials (e.g. a particular asteroid is either detected or not). The number of trials needed to measure a quantity to 95\% confidence is determined by \begin{equation}
    \hat{p} \pm z\sqrt{\frac{\hat{p}(1-\hat{p})}{n}}
\end{equation} where $\hat{p}$ is the fraction of successes in a Bernoulli trial process with $n$ trials and $z$ is the probit of the target error rate (equal to 1.96 for 95\% confidence). For an expected survey completeness of $\hat{p}=$82\% in 5 years (see Section 6), we require $n\sim$7,000 objects in the population to model survey completeness to $<1$\% at 95\% confidence. If we want to be able to divide the population into subgroups (for example, separating out objects by MOID or dynamical class, or by size), we will want an appropriately larger number of asteroids. To this end, we have created 25 solar system representations (SSRs) each with 25,000 NEAs larger than 140 m to assess survey completeness.
 
\subsubsection{Background Object Model}
The main belt asteroids (MBAs) are the dominant source of background objects for NEO Surveyor; we estimate that MBAs will outnumber NEOs by roughly 1000:1 in every \visit~ near the ecliptic plane. Therefore, it is important to model this background population well enough that tracklet linking efficiency and orbit determination can be evaluated accurately in the presence of such objects. A model of non-NEO background populations was generated that includes Mars crossers, inner, middle, and outer MBAs, Hungarias, and more distant objects. For purposes of this simulation, the ``inner MBAs'' are defined as having perihelion distance $q>$1.3 au, semimajor axis $a\leq$2.5 au, and eccentricity $e<0.95$. ``Middle'' MBAs are defined as $q>$1.3 au, 2.5$<a\leq2.82$ au, and eccentricity $e<0.95$. ``Outer'' MBAs are defined as $q>$1.3 au, 2.82$<a\leq$3.6 au, and eccentricity $e<0.95$.  At the sensitivity levels of NEO Surveyor, the background population is dominated by small Mars crossers and inner MBAs. 

Similar to the method for generating the synthetic NEAs, generation of the background objects begins with selecting the diameters using the inverse transform method to randomly sample the cumulative size distributions of the populations down to a specified minimum size. Like the NEAs, the inner, middle, and outer MBAs' size distributions each follow a triple power law of the form $N > D^{-\alpha}$. Fitting this function to each of the three groups using data from NEOWISE \citep{Mainzer.2019a}, which is based on the fits found in \citet{Masiero.2014a}, results in the slopes and breaks shown in Table \ref{tab:mba_size_dist}. For simplicity, the cumulative size distribution of Mars-crossing asteroids and Hungarias was assumed to be identical to that of the inner MBAs. 

The minimum size for all three populations was determined by assessing the probability that 90\% of objects of a given size would ever be detectable by NEO Surveyor. This was assessed by placing the objects at their perihelia at 120$^{\circ}$ solar elongation (equivalent to the region where the NEO Surveyor 5-$\sigma$ sensitivity is 50 $\mu$Jy). This corresponds to size limits of 400 m, 500 m, and 750 m in the inner, middle, and outer main belt respectively, for a total of 12.2 million objects. While it is possible that there is a break in the size distribution at larger sizes than these limits \citep[c.f.][]{Bottke.2020a}, they represent conservative assumptions about the possible number of background objects that NEO Surveyor could detect for purposes of understanding the potential for confusion with NEOs. In total, we estimate that NEO Surveyor will be highly complete for MBAs: completeness is estimated to be $>$99\% and $>$91\% for objects down to 750 m in the inner and middle belt, respectively. Completeness will likely exceed 90\% in the outer belt for objects larger than 1 km.

\begin{figure}[ht!]
\plotone{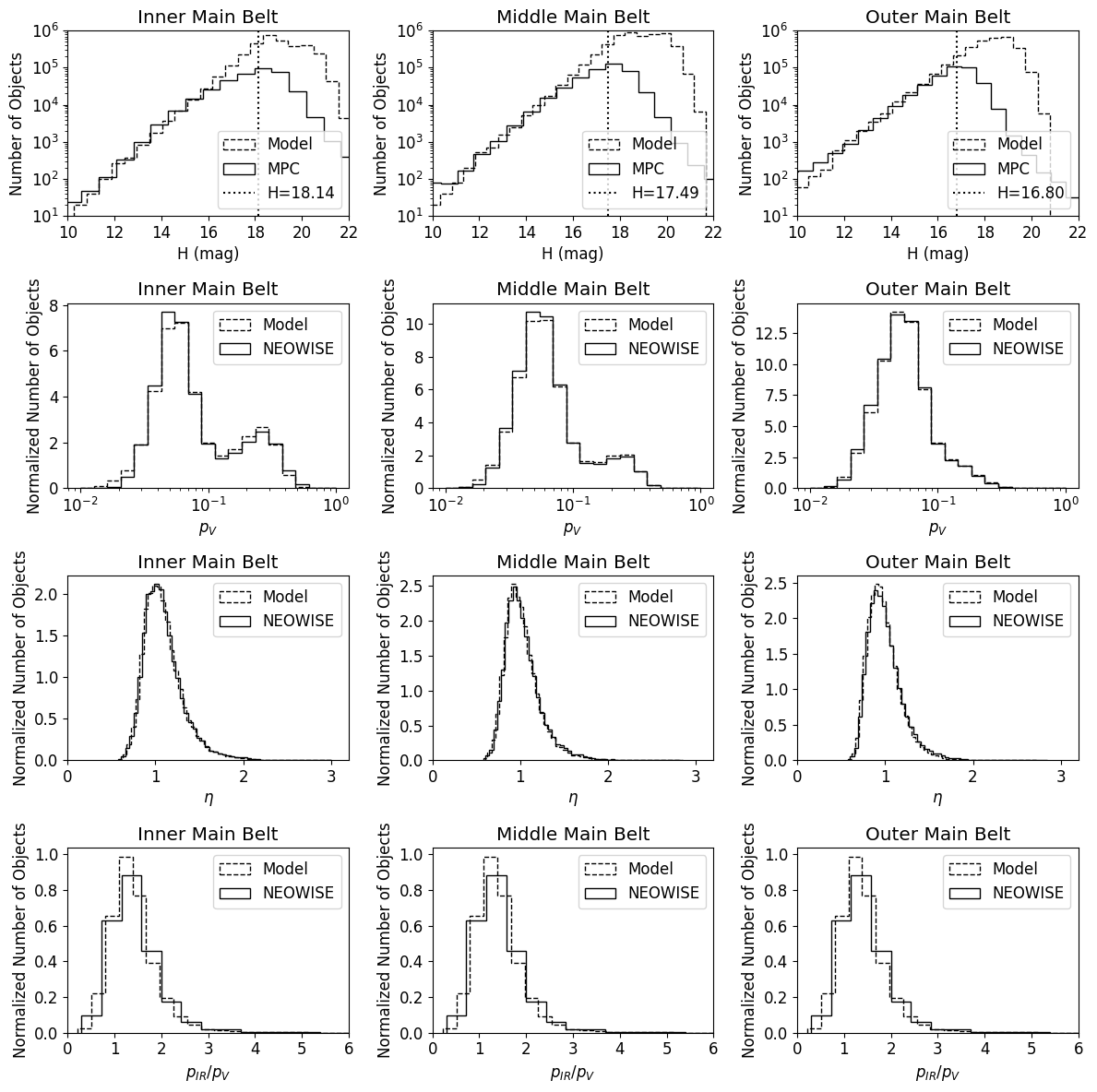}
\caption{Top row: Distribution of absolute visible magnitudes ($H$) for the RSBPM model (dashed lines) for inner, middle, and  outer MBAs compared to known objects in the MPC's holdings (solid lines). The median $H$ magnitude for the MPC catalog for each population is shown as a dotted vertical line. Known objects brighter than $H<$17.75, 17.25, and 16.75 mag respectively were used as the basis for randomly drawing orbital elements for the synthetic objects in the inner, middle, and outer main belt. Second row from top: Normalized visible geometric albedo distribution $p_{V}$ of the model inner, middle, and outer MBAs compared to the measurements of $\sim$130,000 known objects detected based on their 12- or 5-$\mu$m fluxes by NEOWISE. Third row from top: Normalized distribution of beaming parameters ($\eta$) for the RSBPM model compared to the NEOWISE-detected inner, middle, and  outer MBAs for which $\eta$ could be measured. Bottom row: Normalized fraction of $p_{IR} / p_{V}$ for the RSBPM model compared to the NEOWISE-detected inner, middle, and outer MBAs for which $p_{IR}$ could be measured.}
\label{fig:mba_phys_prop}
\end{figure}

Visible geometric albedos were selected for each of the populations using a Gaussian KDE to select from the objects with fitted $p_{V}$ (denoted by fit code ``V'') found in \citet{Mainzer.2019a}. Albedos were limited to 0.015$<p_{V}<$0.70. Beaming parameters were similarly selected using the KDE from the objects for which beaming could be fitted in the NEOWISE data  \citep[fit codes containing ``B'' in][]{Mainzer.2019a}. Albedos at 4.6 $\mu$m were selected by randomly drawing the ratio $p_{IR}/p_{V}$ from the objects in \citet{Mainzer.2019a} with fit codes of ``I'' and multiplying this ratio by the object's $p_{V}$, subject to the condition that 0.015$<p_{IR}<$0.70. Figure \ref{fig:mba_phys_prop} shows the comparison of the model's physical properties elements to data from the MPC and NEOWISE.

Similar to the NEAs, orbital elements were selected from the MPC's sample of objects thought to be observationally complete using a Gaussian KDE. Observational completeness is defined as the turnover point in the $H$ magnitude distribution of known objects as recorded in the MPC's catalog. Figure \ref{fig:mba_phys_prop} (top row) shows the completeness limits for the MBAs as a function of $H$ magnitude, taken to be $H<$17.75, 17.25, and 16.75 mag for the inner, middle, and outer MBAs respectively. 

In order to reduce the data volume of objects in the model, only those asteroids that have a chance of being detectable by NEO Surveyor were saved. As described above, an object was determined to be potentially visible by placing it at its perihelion at 120$^{\circ}$ solar elongation and computing its thermal flux at NC2 using its diameter, $p_{V}$, and $\eta$.  Only objects with NC2 fluxes $>$50 $\mu$Jy are kept. 

\begin{figure}[ht!]
\plotone{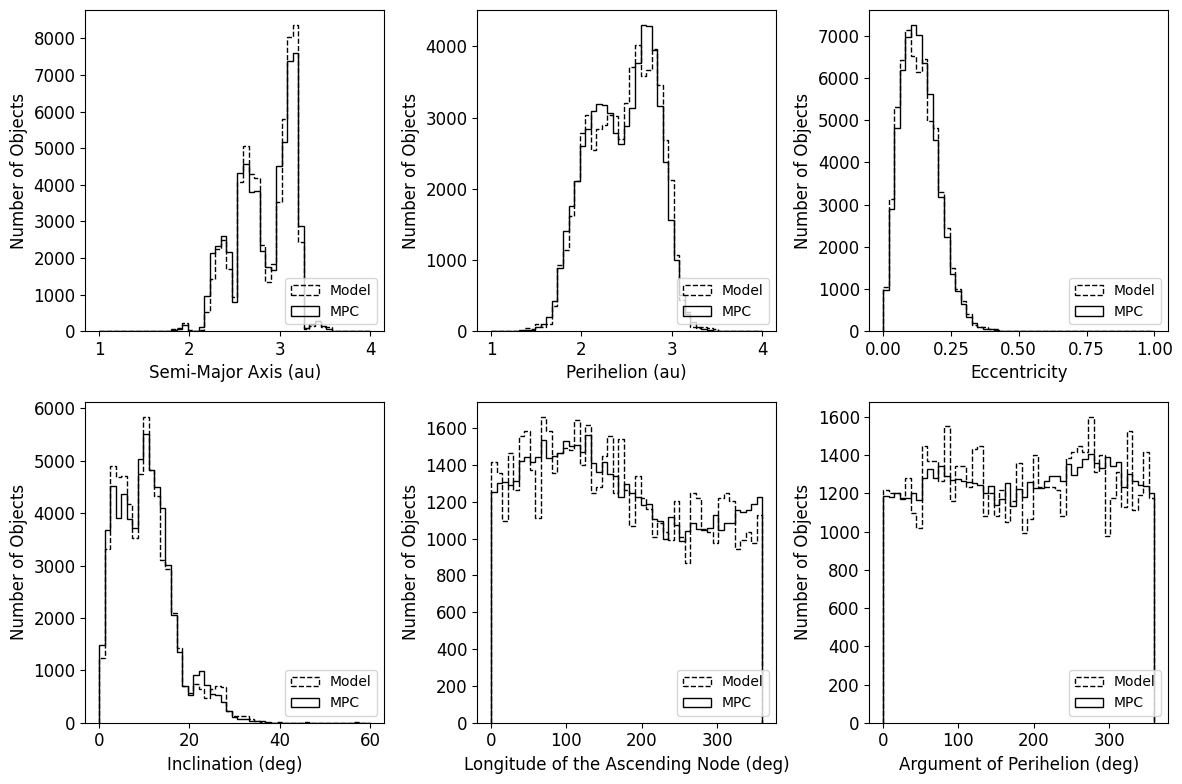}
\caption{Orbital properties of the model background objects compared with those of the known objects. 
\label{fig:mba_orb_props}}
\end{figure}

\begin{figure}[ht!]
\plotone{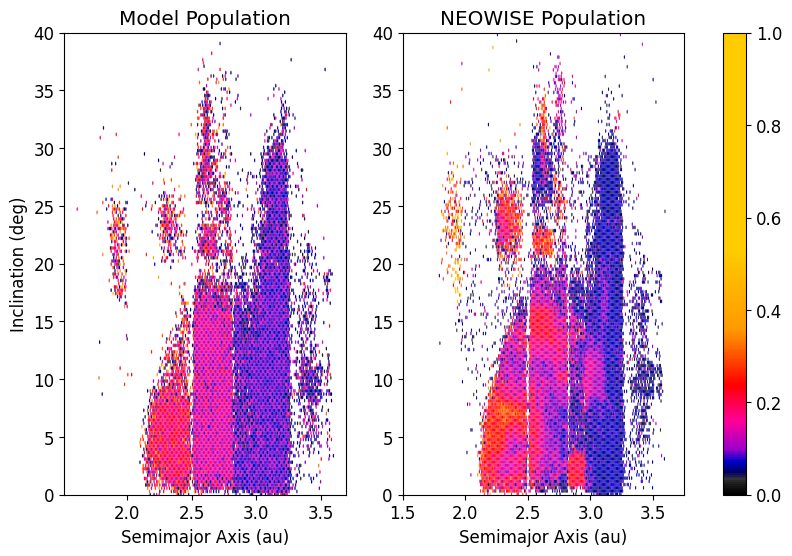}
\caption{Albedo distribution as a function of orbital elements of (left) the reference model population and (right) NEOWISE-detected objects from \citet{Masiero.2013a}.
\label{fig:mba_orb_albedo}}
\end{figure}

Figure \ref{fig:mba_orb_props} shows the comparison of the model's orbital elements to data from the MPC and NEOWISE for objects with $H<$15 mag (the population that is observationally complete throughout the main belt). Figure \ref{fig:mba_orb_albedo} shows the comparison of the model to NEOWISE data when $p_{V}$ is taken into account; while the model lacks the defined albedo structure of collisional families \citep{Masiero.2014a}, it faithfully reproduces the overall gradient of darkening albedo with increasing semimajor axis and produces an appropriate number of background objects near the NEO Surveyor sensitivity limit.

\begin{deluxetable}{cccccccc}
\label{tab:mba_size_dist}
\tablecaption{Slopes and breaks for the triple power laws used to model the inner, middle, and outer MBAs.}
\setcounter{table}{1}
\tablehead{\colhead{Population} & \colhead{$\alpha_{0}$} & \colhead{$\alpha_{1}$} & \colhead{$\alpha_{2}$} & \colhead{$D_{0}$} & \colhead{$D_{1}$} & \colhead{$N_{1km}$} & \colhead{$D_{max}$} \\ 
\colhead{} & \colhead{} & \colhead{} & \colhead{} & \colhead{(km)} & \colhead{(km)} & \colhead{} & \colhead{(km)} } 
\startdata
Inner MBAs &  2.83855 &  1.06297 &  2.08754 &  12.9908 &  59.6530 &  245917 & 196.371 \\
Middle MBAs &  2.80428 &  1.21062 &  3.96481 &  12.4747 &  103.0874 &  804209.5 &  231.689 \\
Outer MBAs &  2.47882 &  1.34556 &  2.6616 &  19.276 &  75.342 &  1600322 &  453.239 \\
\enddata
\end{deluxetable}

\subsection{Survey Plan Generation}
The simulator includes a survey plan modeling tool (SPMT) that during mission development generates the survey plan, which consists of a sequence of \emph{Visits} which respect the Observatory's operable orientation zones. The SPMT creates the successive \emph{Loops} (which produce \emph{Quads} of detections that are linked together to form tracklets of minor planet detections), \emph{Stacks}, and \emph{Sides} for the duration of the 5-year baseline mission; see Figure \ref{fig:survey_pattern} for a representation of the survey pattern. 

The SPMT first creates a grid of positions in latitude and longitude that are fixed on the sky, then uses the ephemeris of the telescope in its SEL1 halo orbit to compute the line of sight with respect to the predicted positions of the Sun and Moon. The list of pointings is recursively edited to avoid keep out zones around the Moon and the Sun. Nominally, the survey will maintain a minimum separation of 20$^\circ$ from the Moon to minimize the impacts of scattered moonlight. \emph{Loops} with fewer than six \emph{Visits} are deleted to ensure that tracklets span at least one hour.

The SPMT also edits the list of pointings to avoid taking \emph{Visits} during specified times when the telescope is unavailable due to downlinks, slews, momentum management, calibrations, and orbit station-keeping maneuvers. During mission development, these ``survey keep-out windows" (SKOWs) are scheduled algorithmically for the entire mission in advance of SPMT runs by an adaptation of a high-heritage spacecraft activity planning framework that has been used in the past to generate Mars 2020 cruise background sequences, schedule Mars 2020 relay orbiter communication, and create command products for the Mars \emph{InSight} lander. The SKOW events are rescheduled periodically as the concept of operations matures to maximize time available for survey. After the SPMT runs, branching off from the data flow of the survey performance simulator described below, the same program that creates the SKOW events reads back in the SPMT-generated survey plan to create an integrated set of mission timelines and orientation plans. These derived products are distributed to the wider NEO Surveyor team, which feeds and improves consistency between various investigation and engineering analyses.

\subsection{Ephemeris Propagation}
To eliminate the need to store large files of synthetic object state vectors, an N-body orbit state propagation code is used to compute the positions of the synthetic objects in the simulation. The code uses an initial value problem formulation with a numerical integrator (scipy's \emph{solve\_ivp} function using the `LSODA' method), taking the initial positions and velocities of each object and the major sources of perturbations (the planets and the largest asteroids) and computing their subsequent positions and velocities using the sum of the forces from the N bodies acting on them. In the initial value problem formulation (IVP) method, the IVP can solve for $\frac{df}{dt} = y(f(t))$, provided the initial condition $f(t_{0})$ is known. In the case of orbit propagation, the initial positions and velocities of all planets and bodies are known, and their derivatives with respect to time are their velocities and accelerations: \begin{equation}
    \frac{\delta[p_{0}, p_{1}, ...v_{0}, v_{1}, ...]}{\delta t} = [v_{0}, v_{1}, ..., a_{0}, a_{1}, ...] 
\end{equation}  
where $[p_{0}, p_{1}, ...v_{0}, v_{1}, ...]$ are the positions and velocities of all perturbing bodies and the particles respectively, and $[a_{0}, a_{1}, ...]$ are the accelerations; these are simply the sum of all forces acting on each object. The code takes into account the general relativistic force due to the Sun using \begin{equation}
    \mathbf{a}_{REL}=\frac{G M_\odot}{c^{2}|\mathbf{r}|^{3}}\Bigl[\Bigl(\frac{4GM_\odot}{|\mathbf{r}|} - |\mathbf{v}^{2}|\Bigr)\mathbf{r} + 4(\mathbf{r}\cdot\mathbf{v})\mathbf{v}\Bigr]
\end{equation}

\noindent following the method recommended in \citet{Farnocchia.2015a}, where $G$ is the gravitational constant, $c$ is the speed of light, $M_\odot$ is the Sun's mass, $\mathbf{r}$ is the position of the body, and $\mathbf{v}$ is its velocity. These initial conditions and forces acting on the bodies are supplied to the IVP, and it solves for the positions and velocities at the next time interval.

This N-body propagation method is used to compute positions at specified intervals (taken here to be one hour) for all objects in the simulation; positions at the individual \emph{Visits} between these one-hour intervals are computed through linear interpolation to improve computational speed. Comparison of the outputs of the positions computed in this fashion were compared with Horizons positions of known objects, and they agree to within 0.3 arcsec for 99.95\% of \emph{Visits} over the 5-year duration of the survey (see Masiero et al. 2023). We require that computed ephemeris uncertainty for 99.9\% of synthetic objects in the survey simulation be less than 0.4 arcsec, so that the analysis is dominated by the astrometric uncertainty of the positional measurements rather than computational precision.

\subsection{Field of View Checks}
Once the ephemerides for all bodies in the model have been determined for the times of all \emph{Visits} in the survey, a comparison to the NEO Surveyor field of view is made to determine whether or not each object fell within it in each of the two channels. The simulator takes into account positional uncertainty due to the modeled performance of the spacecraft attitude determination and control system and the gaps between the individual detectors in each focal plane module. Both NC1 and NC2 fields of view are required to overlap to within 95\% of their projected areas on the sky to maximize the chances of an object being detected in both bands. The positions and velocities of all potential detections that fell within a chip are stored in a database.

\subsection{Detection}
The next step in the simulation is to assess whether a potential detection (defined as an object that appears within the field of view of an individual chip) would be bright enough to be detected. This requires computation of the flux from the object and comparison to the sensitivity of the survey, which varies across the sky as a function of latitude and longitude. The driving source of background at 8 $\mu$m is the zodiacal background, particularly at the lowest solar elongation that NEO Surveyor can observe (45$^{\circ}$), while at 4.6 $\mu$m it is a combination of detector read noise and zodiacal background. A grid of sensitivities is computed across the sky, taking into account the image quality and throughput of the optical system, detector quantum efficiency, contributions from the natural astrophysical background, and thermal emission from the optical system. 

The model assumes a probabilistic detection model based on NSDS image simulation results in which objects  with signal-to-noise ratios (SNR) between 5-10 are detected 85\% of the time; this increases to 95\% detection probability for SNR=10-20 and 99\% for SNR$>$20. Objects below SNR=5 are not counted as detected. This is likely to be a conservative estimate based on the performance of NEOWISE, which currently forms tracklets out of detections made with SNR$>$4.5 \citep{Mainzer.2014a}\footnote{https://wise2.ipac.caltech.edu/docs/release/neowise/expsup/sec4\_3.html\#Adaptations}. NEO Surveyor uses detections made on differenced images,
rather than direct images, which helps to suppress background sources.

Small body fluxes are modeled using an implementation of the near-Earth asteroid thermal model \citep[NEATM;][]{Harris.1998a}, with an option to use the fast-rotating model \citep[FRM;][]{Lebofsky.1978a}. Objects are modeled as spheres with points placed evenly on them using a variation of the Fibonacci Lattice algorithm. Phase curve parameter $G$ (from the \citet{Bowell.1989a} H-G asteroid phase curve formalism)  was assumed to be 0.15 mag for all objects, and emissivity was set to 0.9. The algorithm was validated through comparisons to the NEOWISE data for NEOs and MBAs. See Masiero et al. 2023 for more details on this validation. 

\begin{figure}[ht!]
\plotone{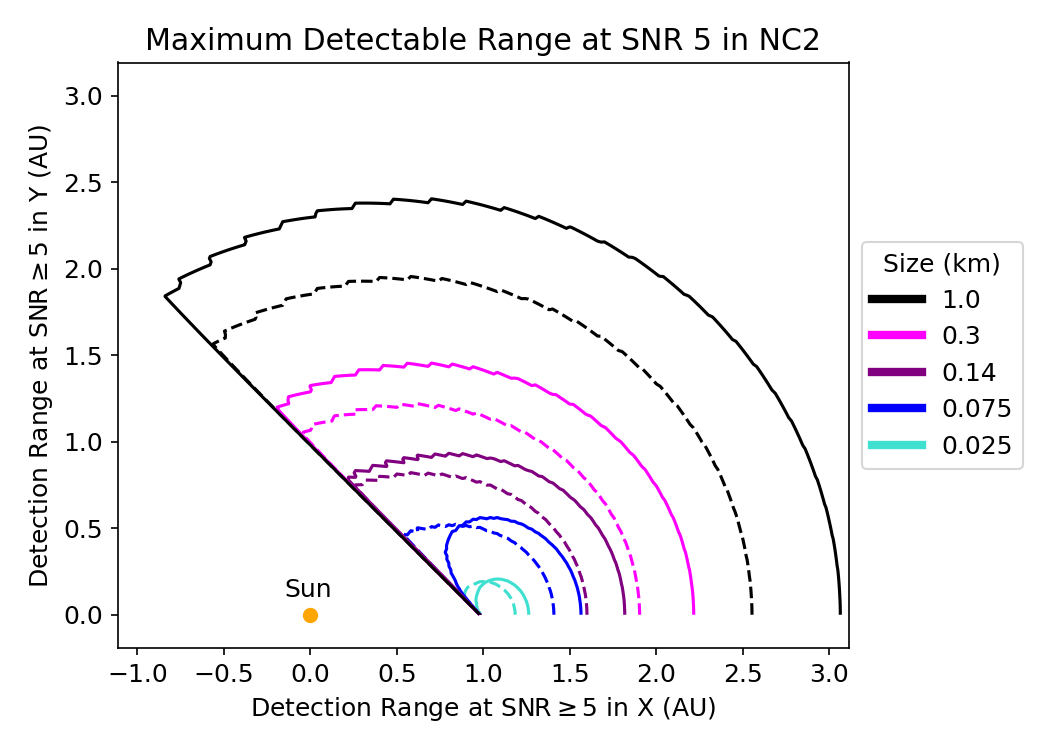}
\caption{Detectable range for NEO Surveyor at SNR=5 or greater of typical NEOs at sizes ranging from 25 m to 1 km with $p_{V}$=0.15 computed using the NEATM model (solid contour lines) and the FRM model (dashed contour lines). Since NEAs below $\sim$200 m are more likely to be rapid rotators \citep{Pravec.2008a}, the FRM model is probably the more appropriate choice for objects in this size range and lower. 
\label{fig:neos_detectability}}
\end{figure}

Figure \ref{fig:neos_detectability} shows the distances at which NEOs of different sizes can be detected for NEO Surveyor, depending on the choice of thermal model (NEATM vs. FRM). 

\subsection{Assembling Tracklets and Tracks}
The penultimate step in the survey simulation is to determine which individual detections of minor planets can be reliably linked into ``tracklets". The NEO Surveyor image simulator is used to assess the efficiency with which tracklets can be linked by creating simulated images implanted with synthetic moving objects and extracting them and linking them together using NMODE. The linking efficiency is supplied to the survey simulator as a single parameter taken to be 99\% based on historical experience from the performance of linking tracklets from the NEOWISE mission to known or new objects. NEOWISE is the closest analog at present, but a robust testing regime is being planned with the MPC to confirm this assumed value. Moving object velocities must fall within $0.008 - 8.000^\circ$/day to be counted as a tracklet; the lower limit is set by the need to be able to distinguish motion in the $\sim2~$hours between \emph{Visits} that observe the same area of sky, using the $\sim$3 arcsec NC2 point spread function. For a bright source, motion of $\sim$0.25 PSFs between \emph{Visits} should be detectable. The upper velocity limit is set by the point at which trailing losses of the source within individual \emph{Exposures} become significant.   

The final step in the simulation of NEO Surveyor's performance is to determine which tracklets can be linked into ``tracks", which are sets of at least two tracklets spanning $\sim$13 days. This step is performed by the MPC. The linking efficiency is modeled as being 99\% based on historical experience from the performance of linking tracklets from the NEOWISE mission to known or new objects (as opposed to being sent to the MPC's Isolated Tracklet File). At present, NEOWISE is the closest analog, but a robust test plan is being carried out with the MPC to confirm this assumed value. Once a track is formed, in most cases an orbit can be computed for the object, leading to it being formally designated as a discovery by the MPC if the object has not been previously identified. At this point, an object is counted by the survey simulator and included in the assessment of survey completeness.

Astrophysical background sources will be removed via image subtraction, so will not significantly impact tracklet building or linking. There will, however, likely be confusion in linking together the MBAs observed, especially at the start of the mission when many will not be previously known. However, because MBAs do not change rapidly in brightness, we expect to see most MBAs every 13 days for long periods of time, which will remove much of the linking ambiguity after the first year or so of operations.

\section{Performance Predictions}
\begin{figure}[ht!]
\plotone{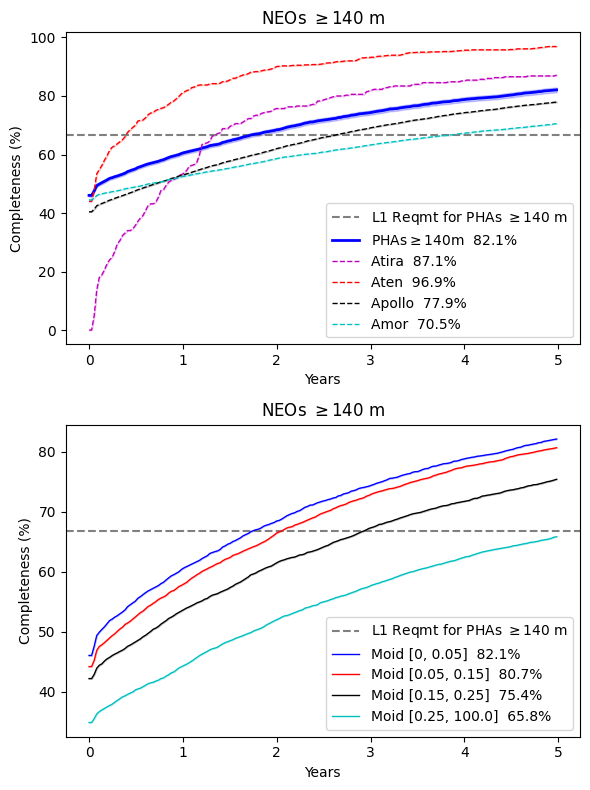}
\caption{Top: Survey completeness as a function of time for potentially hazardous asteroids larger than 140 m in diameter (solid blue line). These simulations show that by the time the mission completes its baseline five year survey, NEO Surveyor will have exceeded the mission's Level 1 requirement of 2/3 survey completeness for PHAs $\geq$140 m in diameter (dashed gray line). The PHA catalog will reach 90\% completeness within 10 years of the launch of Surveyor, fulfilling the Congressional mandate to NASA. The plot does not include possible contributions from Rubin. Survey completeness for Atiras, Atens, Apollos, and Amors larger than 140 m are shown as dashed lines. Bottom: Survey completeness as a function of MOID.
\label{fig:survey_completeness}}
\end{figure}

The NEO Surveyor project builds upon the discoveries made by past and present NEO surveys. To properly account for the objects that they have and will discover between now and launch, a model of the ground-based NEO surveys has been developed (Grav et al. 2023). By fitting a curve to the rate of discovery as a function of sky coverage and sensitivity of the Spacewatch, Catalina Sky Survey and PanSTARRS telescopes, we estimate that the current completeness for NEAs larger than 140 m is $\sim$40\%. The performance of these surveys can be projected into the future, and objects in the synthetic survey are marked as likely to have been found prior to launch. For simplicity, we assumed in the model that all ground-based surveys stopped at the start of the NEO Surveyor survey. The results, including the combined total of the NEO Surveyor survey performance along with the objects that have been or will be found by the existing surveys are shown for objects larger than 140 m in Figure \ref{fig:survey_completeness}. 

\begin{figure}[ht!]
\plotone{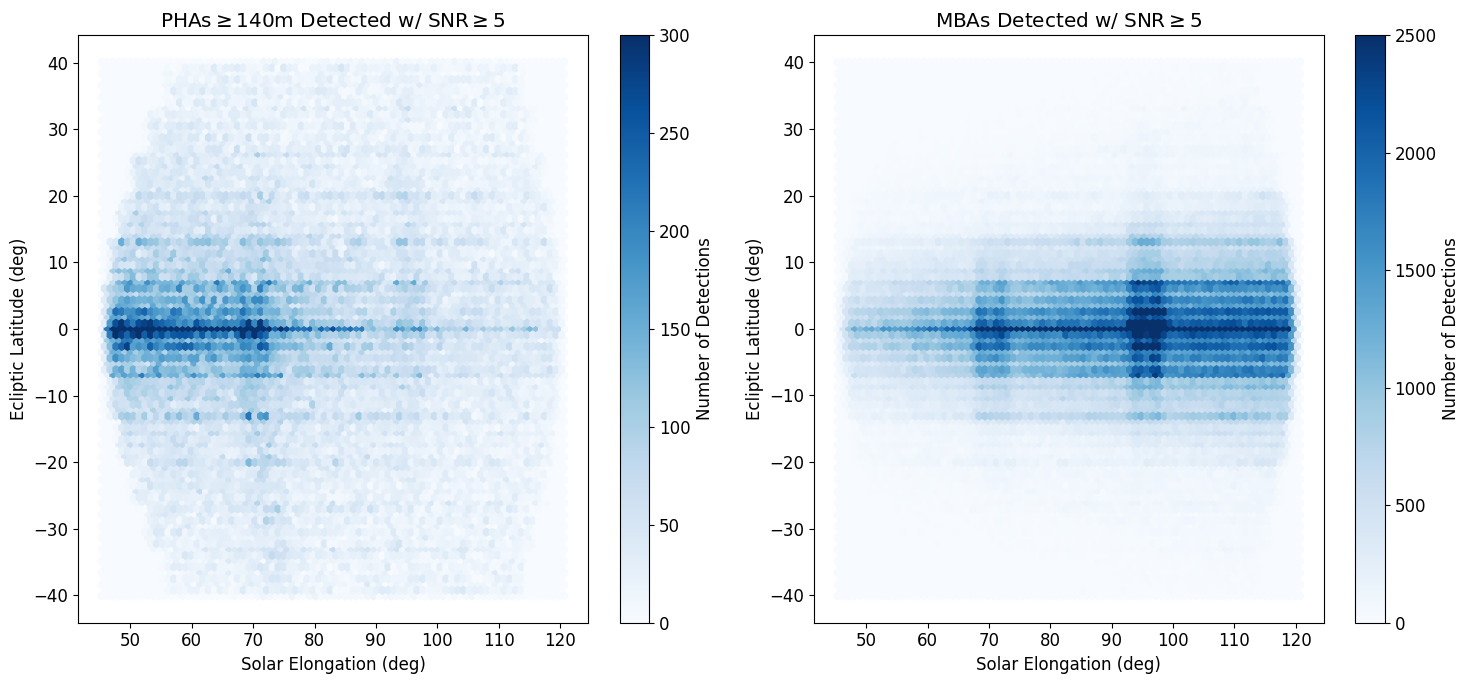}
\caption{(Left) NEO Surveyor detections of PHAs larger than 140 m for five years of observations as a function of sky position. The density of detections is greatest in the direction of the Sun. (Right) Detections of MBAs for five years of observations. In both figures, the bands of enhanced detections are the result of small overlaps between \emph{Stacks}. 
\label{fig:dets_latlon}}
\end{figure}

NEO Surveyor is particularly sensitive to PHAs due to the combination of its sensitivity and its field of regard (Figure \ref{fig:field_of_regard}). NEO Surveyor's view of the regions close to the Sun increases its probability of detecting objects in the most circular, Earth-like orbits, which tend to have lower MOIDs (Figure \ref{fig:survey_completeness}). NEO Surveyor is well-suited to detecting Atens, given their circular orbits, complementing ground-based surveys that search near opposition, which will tend to be more sensitive to Apollos near aphelion and Amors. Figure \ref{fig:dets_latlon} (left) shows the distribution of detections of PHAs as a function of sky position for NEO Surveyor; the density of detections is highest at the lowest solar elongations, illustrating the utility of surveying in these near-Sun regions. This can be compared with the distribution of MBAs (Figure \ref{fig:dets_latlon}, right), which are preferentially distributed at higher elongations on the ecliptic.

We have not attempted to incorporate a model of the Vera C. Rubin Observatory \citep{Ivezic.2019a} in the performance predictions of NEO Surveyor. Rubin collects observations on a different cadence from NEO Surveyor, and its science objectives include an array of other topics. The number of NEOs it will discover depends on the  details of its observational cadence and tracklet/track linking efficiency \citep{Grav.2016a, Veres.2017a, Jones.2018a}. If Rubin discovers many NEOs prior to the NEO Surveyor launch, it will benefit the current goal of planetary defense, which is to come as close as possible to finding and cataloging $>$90\% of PHAs $>$140 m. In addition, we seek to obtain basic physical characterization of these objects and find as many smaller NEOs and comets as possible.  Rubin and NEO Surveyor are thus highly complementary, and both projects are likely to be necessary to achieve these difficult objectives. 

Rubin should help significantly in reducing a major source of confusion, main belt asteroids, as it  should discover and catalog a large number of distant objects that can cause broken tracklets or false linkages. In addition, we seek to obtain visible geometric albedos ($p_{V}$) for as many objects as possible, as albedo has links to composition and taxonomy \citep{Stuart.2004a, Mainzer.2012a}. Rubin will provide measurements of phase curves and visible magnitudes, which in turn will result in well-constrained albedos when combined with thermal measurements from NEO Surveyor. The combination of Rubin and NEO Surveyor data should allow for derivation of albedos for millions of asteroids, providing a significant improvement in our understanding of the distributions of collisional family members as well as NEO origins.

\begin{figure}[ht!]
\plotone{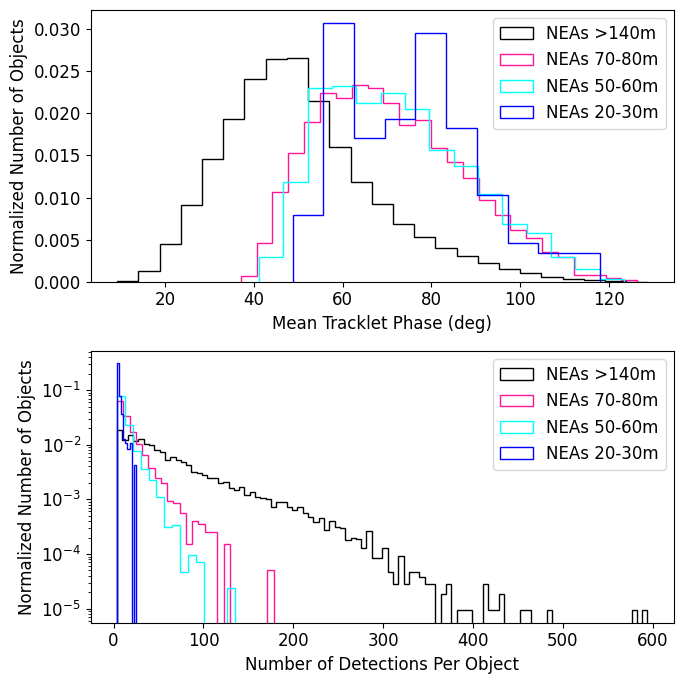}
\caption{Top: Distribution of phase angles for NEAs detected over the 5-year baseline mission, grouped by size. Bottom: Number of detections per NEA as a function of size.
\label{fig:phase_det}}
\end{figure}

Because they can be seen at a greater range of distances, larger objects receive many detections spanning a wide range of phase angles. Their mean observational phase is lower than that of smaller objects, which tend to  be seen when they are closer to the observatory. Figure \ref{fig:phase_det} (top) shows that the phase angles for most objects larger than 140 m are typically less than 60$^{\circ}$, facilitating thermal modeling with NEATM \citep{Mommert.2018}. Smaller objects are more often observed at higher phase angles. Most objects larger than 140 m are observed with up to dozens or even hundreds of detections (Figure \ref{fig:phase_det}, bottom), supporting the use of more detailed thermophysical models \citep[e.g.][]{Delbo.2007a, Delbo.2009a, Ali-Lagoa.2014a, Koren.2015a, Hung.2022a, Satpathy.2022a}.

\begin{figure}[ht!]
\plotone{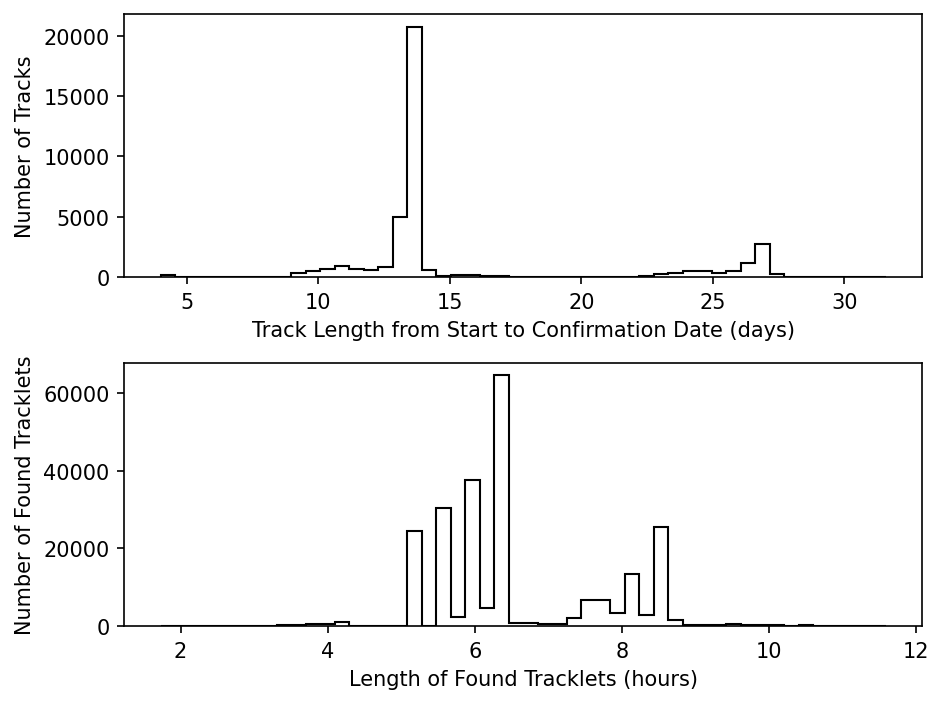}
\caption{Top: Distribution of track lengths for NEAs $\geq$140 m found by NEO Surveyor. Bottom: Distribution of lengths of individual tracklets for these objects.
\label{fig:tracklets_cbe}}
\end{figure}

\begin{figure}[ht!]
\plotone{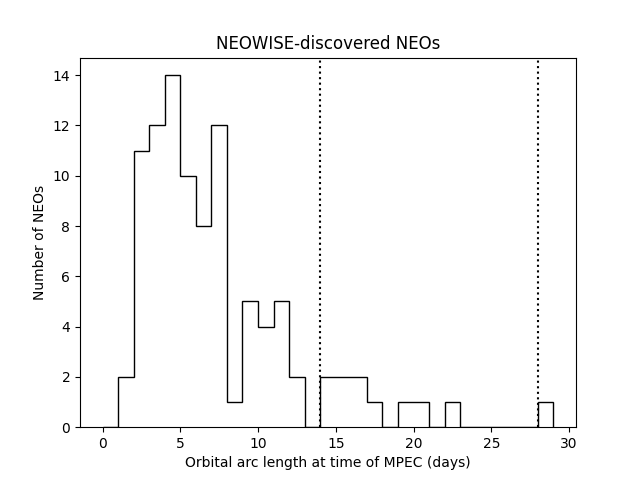}
\caption{Arc length of NEOs discovered by the NEOWISE mission at the time that the MPC issued a Minor Planet Electronic Circular, which we use as a proxy for the object having sufficient data to constrain an orbit. Vertical dotted lines indicate 14 and 28 days. In total, 89\% of the NEOs discovered by NEOWISE had arcs less than 14 days, while 99\% had arcs less than 28 days.
\label{fig:arcs}}
\end{figure}

The default NEO Surveyor cadence is designed for self follow up; objects should not in general become lost. The NEO Surveyor survey cadence results in observational arcs that span more than 10 days for the majority of $>$140 m objects after two years of surveying (Figure \ref{fig:tracklets_cbe}). Most large objects are observed on multiple epochs spanning months to years. This is consistent or better than the distribution of observational arc lengths currently available in the existing catalog of known NEOs discovered by NEOWISE (Figure \ref{fig:arcs}), which observes at a similar geometry. Figure \ref{fig:tracklets_cbe} shows the distribution of tracklet and track lengths for tracklets corresponding to objects that form tracks and are therefore counted as found. The minimum track length is set by the minimum observational arc needed by the MPC to confirm and designate an orbit for a new discovery, based on experience with NEOWISE (Figure \ref{fig:arcs}). Similarly, the minimum length of individual tracklets within each track is set by the desire to have an observational arc long enough to ensure that a tracklet can be successfully linked to a second tracklet for an object up to $\sim$28 days later. The cadence should ease the burden of follow up so that resources can be reserved for objects of unusual interest or to obtain characterization measurements such as spectroscopy, optical phase curves, or more detailed light curves.

\section{Conclusion}
NEO Surveyor will make significant and rapid progress toward the objective of finding and cataloging $>$90\% of PHAs larger than 140 m, in addition to providing measurements of their sizes and in many cases albedos. The results of the mission will significantly improve our ability to mount a successful NEO deflection campaign, should it become necessary. NEO Surveyor will additionally provide diameters for each cataloged object, and will generate a survey that is size-limited survey as opposed to $H$-limited. Its thermal wavelengths will make it especially effective at discovering dark C-type NEOs, which constitute a not-insignificant fraction of the NEO population. 

The impact frequency from smaller NEOs will be determined by computing the size and orbital element distributions of the objects (after accounting for survey biases), then taking these distributions and propagating them forward and backward over the next several thousand years using a suitable numerical integrator to determine the average timescale between impacts. By determining the size-frequency distribution for small NEOs using diameters instead of $H$, it will be possible to significantly improve our understanding of the frequency of Earth impacts in the $\sim$20-140 m size range.

Moreover, by providing two-channel thermal infrared data on millions of main belt asteroids, thousands of comets, and hundreds of thousands of NEOs, NEO Surveyor will support a host of scientific studies of these bodies. Detailed probes of the size-frequency distribution of these populations will help to improve constraints on their origins as well as impact probabilities. In addition, the project will complement multi-wavelength optical data expected to become available from the Rubin Observatory. The publicly-released and archived data products of the NEO Surveyor mission will enable not only the discovery of objects hazardous to the Earth, but also studies of objects both known and yet-to-be found. It will also generate a vast catalog of images of the sky at thermal wavelengths spanning multiple epochs that will support a wide array of astrophysical analyses. In the process of determining the probability of impacts from NEOs in the next century, we will learn a great deal more about the solar system's contents and origins.

\section{Acknowledgements}
This publication makes use of data products from the NEO Surveyor, which is a joint project of the University of Arizona and the Jet Propulsion Laboratory/California Institute of Technology, funded by the National Aeronautics and Space Administration. We gratefully acknowledge the services and support of the International Astronomical Union's Minor Planet Center. This publication makes use of data products from the Wide-field Infrared Survey Explorer, which is a joint project of the University of California, Los Angeles, and the Jet Propulsion Laboratory/California Institute of Technology, funded by the National Aeronautics and Space Administration. This publication also makes use of data products from NEOWISE, which is a project of the Jet Propulsion Laboratory/University of Arizona, funded by the Planetary Science Division of the National Aeronautics and Space Administration. This research has made use of the NASA/IPAC Infrared Science Archive, which is funded by the National Aeronautics and Space Administration and operated by the California Institute of Technology.

We are extremely grateful to our anonymous referees; their comments have materially improved the clarity and quality of this manuscript.

\vspace{5mm}
\facilities{WISE, NEOWISE, IRSA, MPC}

Dataset usage:

\dataset[WISE All-Sky 4-band Single-Exposure Images]{https://www.ipac.caltech.edu/doi/irsa/10.26131/IRSA152}

\dataset[AllWISE Source Catalog]
{https://www.ipac.caltech.edu/doi/irsa/10.26131/IRSA1}

\dataset[NEOWISE-R Single Exposure (L1b) Source Table]
{https://www.ipac.caltech.edu/doi/irsa/10.26131/IRSA144}

\software{astropy \citep{Astropy.2022a}, scipy \citep{Virtanen.2020a}, numpy \citep{Numpy.2020a}, SPICE \citep{Spice.1996a}, spiceypy \citep{Spiceypy.2020a}, Blackbird \citep{Blackbird.2020a}}

\bibliography{references}{}
\bibliographystyle{aasjournal}

\end{document}